\documentclass[twocolumn]{aastex631}
\usepackage{appendix}
\usepackage{isomath}
\newcommand{\Bra }{Br~${\alpha}$}

\newcommand{\Hii}{H{\sc ii}}

\newcommand{\HII}{H{\sc ii}}

\begin{document}
\received{November 21, 2025}
\revised{February 5, 2026}
\accepted{\today}

\submitjournal{ApJ}

\shorttitle{Magnetized HII Regions}
\shortauthors{Crowe et al.}

\title{MHD Simulations of Strongly Magnetized \Hii~Region Evolution: Evidence for Ionized Gas Filamentation}
%Are HII Regions Magnetically Dominated in the CMZ?} I suggest this title change (I don't think phrasing as a question is needed) but we can discuss -stc

% \title{Magnetically Dominated HII Regions in the CMZ:  Sgr C}

\correspondingauthor{Samuel Crowe}
\email{exet6431@ox.ac.uk}

\author[0009-0005-0394-3754]{Samuel Crowe} 

\affiliation{Oxford Centre for the History of Science, Medicine, and Technology, University of Oxford, Oxford OX2 6AH, United Kingdom}
\affiliation{Dept. of Astronomy, University of Virginia, Charlottesville, Virginia 22904, USA} 
\affiliation{Virginia Institute of Theoretical Astronomy, University of Virginia, Charlottesville, VA 22904, USA}

\author[0000-0003-2929-1502]{Yisheng Tu}
\affiliation{Dept. of Astronomy, University of Virginia, Charlottesville, Virginia 22904, USA}
\affiliation{Virginia Institute of Theoretical Astronomy, University of Virginia, Charlottesville, VA 22904, USA}
\affiliation{Department of Astronomy, University of Michigan, Ann Arbor, MI 48103, USA}

\author[0000-0002-7402-6487]{Zhi-Yun Li}
\affiliation{Dept. of Astronomy, University of Virginia, Charlottesville, Virginia 22904, USA}
\affiliation{Virginia Institute of Theoretical Astronomy, University of Virginia, Charlottesville, VA 22904, USA}

\author[0000-0001-6228-8634]{Jeong-Gyu Kim}
\affiliation{Korea Institute of Advanced Study, 85 Hoegi-ro, Dongdaemun-gu, Seoul 02455, Republic of Korea}

\author[0000-0001-8135-6612]{John Bally} 

\affiliation{Center for Astrophysics and Space Astronomy, 
     Department of Astrophysical and Planetary Sciences \\
     University of Colorado, Boulder, CO 80389, USA} 

% List of institutions

\begin{abstract} Recent JWST observations of \Hii~regions in the Central Molecular Zone have shown a highly filamentary morphology distinct from \Hii~regions 
%in the solar neighborhood.
in other parts of the galaxy. We present magnetohydrodynamic (MHD) simulations
%, using the \texttt{Athena++} simulation framework, 
of strongly magnetized (plasma-$\beta\ll 1$) \Hii~region evolution that 
%effectively recreate
investigate and describe the formation of these ionized gas filaments. \Hii~region evolution has been simulated in a $30$ pc$^3$ box, in distinct models with pre-placed overdensities in the ambient medium and overdensities that have been generated with driven turbulence. We find that when these overdensities are seeded in the ambient medium before the birth of the ionizing source, the photoionized plasma stripped off of these dense blobs is funneled into long filaments along the magnetic field lines. The length and emission measure of these ionized gas filaments are similar to the filaments observed in the Central Molecular Zone. Given that these filaments are effectively magnetically-confined flows of photoionized gas, their density and curvature are influenced by the density of the blob and the geometry of the configuration.
\end{abstract}


\section{Introduction}\label{sec:introduction}

    \Hii\:regions, the regions of warm, ionized gas that are typically associated with star-forming molecular clouds, stand at an important intersection in contemporary astrophysics. 
    %In their earliest, compact phases, 
    \Hii\:regions 
    %provide an indelible contribution to the internal stellar feedback that disperses 
    terminate star formation in molecular clouds before the conversion of gas to newborn stars is complete % \citep{zuckerman74,krumholz07}. 
    \citep{zuckerman74,chevance23}. 
    As signposts of star formation, particularly massive star formation, which itself has an outsized impact on the evolution of galaxies \citep[see][for recent reviews]{tan14,rosen20}, \Hii\:regions are crucial diagnostics of star formation rates in the Milky Way and other galaxies \citep{kennicutt89,williams97,debuizer19}. Likewise, % the containment of ionizing radiation to
    the escape of UV radiation from 
    the \Hii\:region of a young OB association
    % (i.e., the escape fraction)
    governs the interstellar and intergalactic radiation field \citep[e.g.,][]{gnedin22, linzer24}, the ionization of the diffuse interstellar medium \citep[e.g.,][]{mccallum24}, and the observed dispersion measure of pulsars throughout the galaxy \citep{cordes02}, among other important astrophysical processes \citep[see also][and references therein]{mckee97, haffner2009}.
    
    Only recently, however, has the evolution of \Hii\:regions, and their impact on star formation, been considered in more extreme regions. The Central Molecular Zone (CMZ), the central few hundred parsecs of the Milky Way \citep{henshaw23}, is one such region, characterized by molecular cloud densities, temperatures, turbulence, and magnetic field strengths orders of magnitude above those typically observed in the Galactic disk \citep[see, e.g.,][]{Bally87,Bally88,giveon02,ferriere09,kruijssen14,ginsburg16}. 
    
    Recent James Webb Space Telescope (JWST) observations of the CMZ massive star-forming region Sagittarius C (Sgr C), at a galactocentric radius (in projection) of $\sim75$pc, have revealed a striking filamentary morphology in the Sgr C \Hii\:region in the \Bra\:hydrogen recombination line \citep{crowe25}. \citet{bally25} suggested several possible origins for the observed filaments, including supernova remnants, photo-dissociation regions, and fossil outflow lobes, but presented compelling evidence that a significant population of the filaments are sculpted by magnetic fields, particularly the large-scale poloidal (i.e., perpendicular to the galactic plane) field in the CMZ.
    
    Magnetic fields play an important role in the evolution of star-forming regions \citep{Hennebelle2019,Pattle2019,Pattle2023}, and several papers have probed the effects of magnetic fields on \Hii\:region evolution in particular. %\citet{krumholz07} introduced a magneto-hydrodynamic (MHD) code and incorporated it with existing hydrodynamic frameworks for \Hii\:region modeling. \citet{arthur11} and \citet{zamoraaviles19}
    \citet{krumholz07} and \citet{arthur11} found that magnetic fields inhibit \Hii\:region expansion perpendicular, but not parallel, to the field lines, producing elongated ellipsoid (rather than spherical) \Hii\:regions. \citet{gendelev12} extend this treatment to ``blister'' \Hii\:regions, configurations where an ionizing source is positioned between 
    a high- and low-density medium (e.g., at the edge of a dense molecular cloud), finding that the \Hii\:region is still elongated along the magnetic field lines, and that the inclusion of magnetic effects may increase the injection rate of energy into the cloud.
    
    Finally, \citet{mackey11} present MHD simulations of ``pillars'' of neutral gas at \Hii\ region edges under uniform B-fields up to $160 ~ \mu G$ (plasma-$\beta\sim0.01$). When strong B-fields are included perpendicular to the pillar axis, elongated ``ribbons'' of ionized plasma were observed expanding perpendicular to the pillar axis (parallel to the B-field). Fields parallel to the pillar axis tended to generate ``ribbons'' of ionized gas extending back into the \Hii\ region, parallel to the pillar axis. The authors suggest that in highly magnetized \Hii\ regions, such filamentary structures would be readily observable in plasma tracers, such as the \Bra\:line.
    
    Most literature on \Hii\ region evolution in a magnetized medium considers relatively weak ($\leq 50 ~ \mu$G) fields typical of Galactic disk star forming regions \citep[see, e.g.,][]{vaneck11}. The magnetic fields in the CMZ are orders of magnitude stronger than in the Galactic disk near the Sun, with CMZ field strengths ranging from 100 $\mu $G to over 1~mG in dense molecular clouds \citep{ferriere09}. Therefore, it may be expected that magnetic fields have an outsized effect on the evolution of \Hii\: regions in highly-magnetized environments like the CMZ, potentially producing the filamentation observed in Sgr C.
    
    In this paper, we present MHD simulations of highly-magnetized \Hii\:region evolution in %CMZ-like 
    %highly-magnetized 
    conditions that attempt to investigate and replicate the formation mechanism of the ionized gas filaments observed in Sgr C. The paper is organized as follows. The simulation setup is described in \S\ref{sec:methods}. The results are shown and discussed in \S\ref{sec:results} and \S\ref{sec:discussion}, respectively. A conclusion is made in \S\ref{sec:conclusion}.

\section{Methods}\label{sec:methods}
    We present a suite of radiation magnetohydrodynamic (RMHD) simulations of \Hii\:region evolution using the grid-based code \texttt{Athena++} \citep{stone20} with additional physics modules for adaptive ray-tracing and photochemistry. The additional radiation modules are largely similar to those used in \cite{kim17,kim18,kim19}.
    %\citet{kim21}, which is an augmented version of the setup used in \citet{kim17,kim18,kim19}. 
    %{\color{red} [Sam: has Kim ever used the modules in another paper? If so cite that one and say our module inheres from that one.]}
    %with the addition of adaptive mesh refinement (AMR), used in select simulation runs.
    
    The evolution of a radiatively and magnetically dominated \Hii\:region is governed by the following set of equations:
    
    \begin{equation}\label{eq:continuity}
        \frac{\partial\rho}{\partial t}+\nabla\cdot(\rho \mathbfit{v})=0,
    \end{equation}
    
    \begin{equation}\label{eq:euler}
        \frac{\partial(\rho \mathbfit{v})}{\partial t}+\nabla\cdot[\rho \mathbfit{vv}+P^*\mathbb{I}-\frac{\mathbfit{BB}}{4\pi}]=\mathbfit{f}_\mathrm{rad},
    \end{equation}
    
    % \begin{equation}
    %     \frac{\partial E}{\partial t}+\nabla\cdot[(E+P^*)\mathbfit{v}-\frac{\mathbfit{B}(\mathbfit{B\cdot v})}{4\pi}]=\mathcal{G}-\mathcal{L}+\mathbfit{v}\cdot \mathbfit{f}_\mathrm{rad},
    % \end{equation}
    
    \begin{equation}\label{eq:induction}
        \frac{\partial \mathbfit{B}}{\partial t}-\nabla\times(\mathbfit{v\times B)}=0,
    \end{equation}
    
    % \begin{equation}{\label{eq:continuity_nonH}}
    %     \frac{\partial n_s}{\partial t}+\nabla\cdot(n_s \mathbfit{v})=n_H\mathcal{C}_s,
    % \end{equation}
    \begin{equation}{\label{eq:continuity_nonH}}
        \frac{\partial n_{\rm H^0}}{\partial t}+\nabla\cdot(n_{\rm H^0} \mathbfit{v})=\mathcal{R}-\mathcal{I},
    \end{equation}

    % \begin{equation}\label{eq:radiation}
    %     \frac{\partial \mathbfit{I}}{\partial t} + c\mathbfit{n}\cdot\nabla \mathbfit{I} = \mathbfit{S}(\mathbfit{I}, \mathbfit{n}),
    % \end{equation}
    \begin{equation}\label{eq:radiation}
        \mathbfit{n}\cdot\nabla \mathbfit{I} = -\chi I
    \end{equation}
    where $P^*$ is the total pressure, including the gas thermal pressure and magnetic pressure, and $\mathbfit{f}_\mathrm{rad}$ is the radiative force per unit volume. 
    % To represent the heating ($\mathcal{G}$) and cooling ($\mathcal{L}$) due to radiation, we adopt a simple temperature prescription, 
    Instead of solving the energy equation with explicit heating and cooling terms, we adopt a simple two-temperature isothermal equation of state, used by \citet{kim17}, where temperature is assigned as a smoothly varying function of the fraction of neutral gas, i.e.
    \begin{equation}\label{eq:temp_prescription}
        T=T_\mathrm{ion}-\Big(\frac{x_\mathrm{n}}{2-x_\mathrm{n}}\Big)(T_\mathrm{ion}-T_\mathrm{neu})
    \end{equation}
    where $x_\mathrm{n}=n_\mathrm{H^0}/n_\mathrm{H}$ is the netural gas fraction and $T_\mathrm{ion}=8000~\mathrm{K}$ and $T_\mathrm{neu}=20~\mathrm{K}$ are the temperatures of the fully ionized and fully neutral gas, respectively. 
    %The gas thermal pressure and energy are subsequently calculated through the adiabatic equation of state. %{\color{blue} [If the temperature is assumed, the equation of state should not be adiabatic; please double check.]}
    Since most dynamics occur in the fully ionized region, we assume the gas is in the ideal MHD limit (Eq. \ref{eq:induction}) throughout the simulation. We ignore the effects of gravity as gravity is not expected to be important in the magnetically and radiatively dominated parsec-scale environment we model. The terms on the right hand side of Eq. \ref{eq:continuity_nonH}, represent creation and destruction of atomic hydrogen due to case-B radiative recombination ($\mathcal{R}$) and photoionization ($\mathcal{I}$), respectively.
    % $n_s=x_s n_H$ refers to the number density of species ``s," where $x_s$ is the fractional abundance relative to hydrogen. We explicitly follow the nonequilibrium abundances of hydrogen in molecular (H$_2$), atomic neutral (H$^0$), and ionized (H$^+$) phases but assume equilibrium abundances for carbon- and oxygen-bearing species (C$^0$, C$^+$, CO, O$^0$). Every species followed is passively advected with the velocity field, with $\mathcal{C_s}$ on the right hand side representing the net creation rate from various procsses \citep[see][for more informaation]{kim21}.
    The remaining symbols have their usual meanings. %{\color{red} [Sam: Please double check if this assumption is justified by calculating the ratio between gravitational energy, gas thermal energy, and magnetic energy. For gravity, you can assume some-100 $M_\odot$ star at the center and calculate the gravitational binding energy, then compare the expected value with gas thermal energy and magnetic energy.]}.

    %I checked this, and the magnetic energy in the simulation was ~3e52 erg, thermal ~4e49 erg, gravitational binding ~1.5e48 erg. So our assumption that gravity is negligible seems accurate 

    Photo-ionization is treated by solving the time-independent radiation transfer equation (Eq. \ref{eq:radiation}) with photon sources representing the ionizing photons emitted by hot stars (such as O stars and B stars) in CMZ. 
    For the absorption coefficient per length, we adopt $\chi = n_{\rm H}\sigma_{\rm d} + n_{\rm H^0}\sigma_{\rm pi}$ with a spatially constant dust cross section of $\sigma_{\rm d} = 1\times10^{-21} \,{\rm cm}^2$ and photoionization cross section of $\sigma_{\rm pi} = 3\times10^{-18}\,{\rm cm}^2$.
    The hydrogen ionization fraction due to radiation is calculated with Eq. \ref{eq:continuity_nonH} using a simplified photochemistry network including photo-ionization and recombination. Because of the high ionizing photon emission rates of the hot stars modeled in the simulation, most gas in the simulation stays 
    %either fully ionized (if inside the ionization front) or fully neutral (if outside the radiation front).
    either fully ionized or fully neutral, with the boundary defined by the ionization front.
    We model a $30$ pc$^3$ region with $256$ cells in each direction (i.e., the resolution is 0.12 pc per cell). An ionizing source was placed at $(x,y,z)=(0,0,0)$ with an ionizing photon emission rate of $Q_0=1\times10^{50}$~s$^{-1}$, roughly corresponding to the typical ionizing photon fluxes of star-forming clouds like Sgr C \citep{simpson18a}. 
    %representing the combination of Wolf-Rayet stars and massive main sequence stars in Sgr C \citep{clark21,nogueras-lara24}. 
    %To model the magnetically-dominated environment in Sgr C, 
    A $\hat{z}$-direction magnetic field of 1 mG was added so that the maximum plasma-$\beta$ is $<1$ in all regions (average $\beta\approx10^{-5}$; in the densest regions, where $n(H)\approx10^3\:\mathrm{H/cm^{-3}}$, $P_T/k_B\approx1\times10^7\:\mathrm{Kcm^{-3}}$, whereas $P_B/k_B\approx3\times10^8\:\mathrm{Kcm^{-3}}$). This extremely low $\beta$ value means that the magnetic field changes very little in magnitude or direction across all of the simulations. %{\color{red}$10^{-3}$} in the simulation. {\color{blue}[initial plasma-beta based on the uniform density before the added density perturbations?]} 
    The outflow boundary condition is applied at all simulation boundaries. 
    %The temperature of the fully neutral gas was set to 20 K, and the temperature of the fully ionized gas was set to 8000 K \textbf{NOTE: I am unsure about the details of how the temperature is treated, and what, if any, active calculation of heating and/or cooling is done. Perhaps Kim can advise on this?}. An outflow boundary condition was used. \textbf{Note: more info needed for boundary condition?}
    
    With the above setup, we present two models: We first illustrate one possible filament-forming mechanism using a model with artificially placed spherical overdensities in the ambient medium (hereafter referred to as ``blobs''). Two blobs, each 1 pc in radius, are placed at $(x,y,z)=(5,0,0)$ and $(x,y,z)=(-5,0,5)$. The density in the blobs is uniform at $1000$ H/cm$^{-3}$. The remaining simulation volume is filled with an ambient medium of density $100$ H/cm$^{-3}$. 
    %In this simulation, adaptive mesh refinement (AMR) was also used, with a threshold for the first level of refinement set at $500$ H/cm$^{-3}$, a factor beteween levels of refinement of $3\times$ (i.e., the density must be $3\times$ higher to reach the second level of refinement, etc.), and 3 total levels of refinement. AMR was used for this simulation since there was a relatively small area of interest (i.e., the region around the blobs). 
    
    In the second model,
    %{\color{red}model [Sam: if you will present more than one model, use ``second set/suite of models'', and change other places accordingly]}, 
    we demonstrate how the mechanism identified in the first model naturally forms filament-like structures in a more heterogeneous
    %realistic 
    initial condition where density inhomogeneities were seeded by turbulence.% rather than artificially placing blobs. 
    To %naturally 
    generate a density inhomogeneity for our initial condition, we drive turbulence in the model for 20 Myrs, with no magnetic field or ionizing source. The turbulence driving time is chosen such that the resulting density contrast is similar to the run with pre-set blobs.
    %reflects conditions typical of molecular clouds \citep{rathborne14}. 
    %that have also been shaped by additional processes such as protostellar winds and outflows {\color{red} [Sam: any citations?]}. {\color{blue}[The origin of the supersonic turbulence in a molecular cloud is complicated, including shearing in the Galactic central potential well, possibly MRI on galactic scale, cloud-cloud collisions, SN feedback, and outflow feedback.] Perhaps just leave out "that have ..."} 
    After 20 Myr, we stop the turbulence and introduce the magnetic field and ionization source. The turbulence is driven with an energy spectrum %similar to the Kolmogorov Cascade {\color{blue}[Kolmogorov has a different slope of 5/3, I believe; perhaps remove "similar to..."]}
    \begin{equation}\label{eq:turbulence}
        E(k) \propto k^{-2} 
    \end{equation}
    where $k=2\pi/\lambda$ is the wavenumber for a given eddy with wavelength $\lambda$. Energy was injected into the system at a rate of $2.23\times10^{34}$ erg~s$^{-1}$, with wave numbers between $2\leq\frac{2\pi}{\lambda}\leq 128$. An initial ambient density of $600$ H/cm$^{-3}$ is used to mitigate the mass loss at the simulation boundary during the driven turbulence phase. The average density of the ambient medium upon activation of the ionizing source was much lower, close to the $100$ H/cm$^{-3}$ used for the blob simulations.
    %AMR was not used for these simulation runs because, since the density was nonuniform, we were interested in the behavior over all parts of the simulation domain.
\section{Results}\label{sec:results}
    \subsection{Simulation with pre-placed blobs}\label{sec:blob_simul}
    \begin{figure*}
        \centering
        \includegraphics[width=0.45\textwidth]{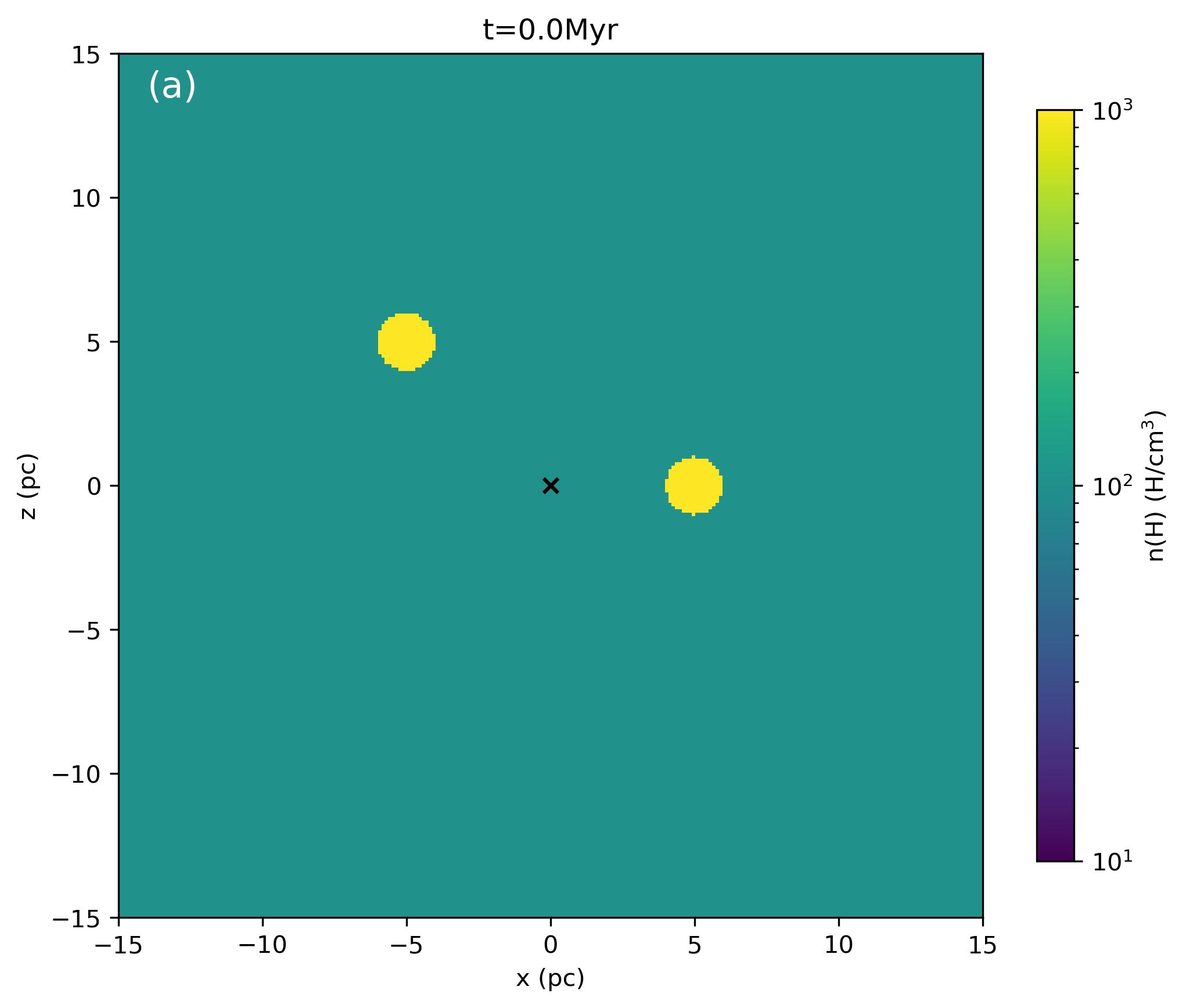} 
        \includegraphics[width=0.45\textwidth]{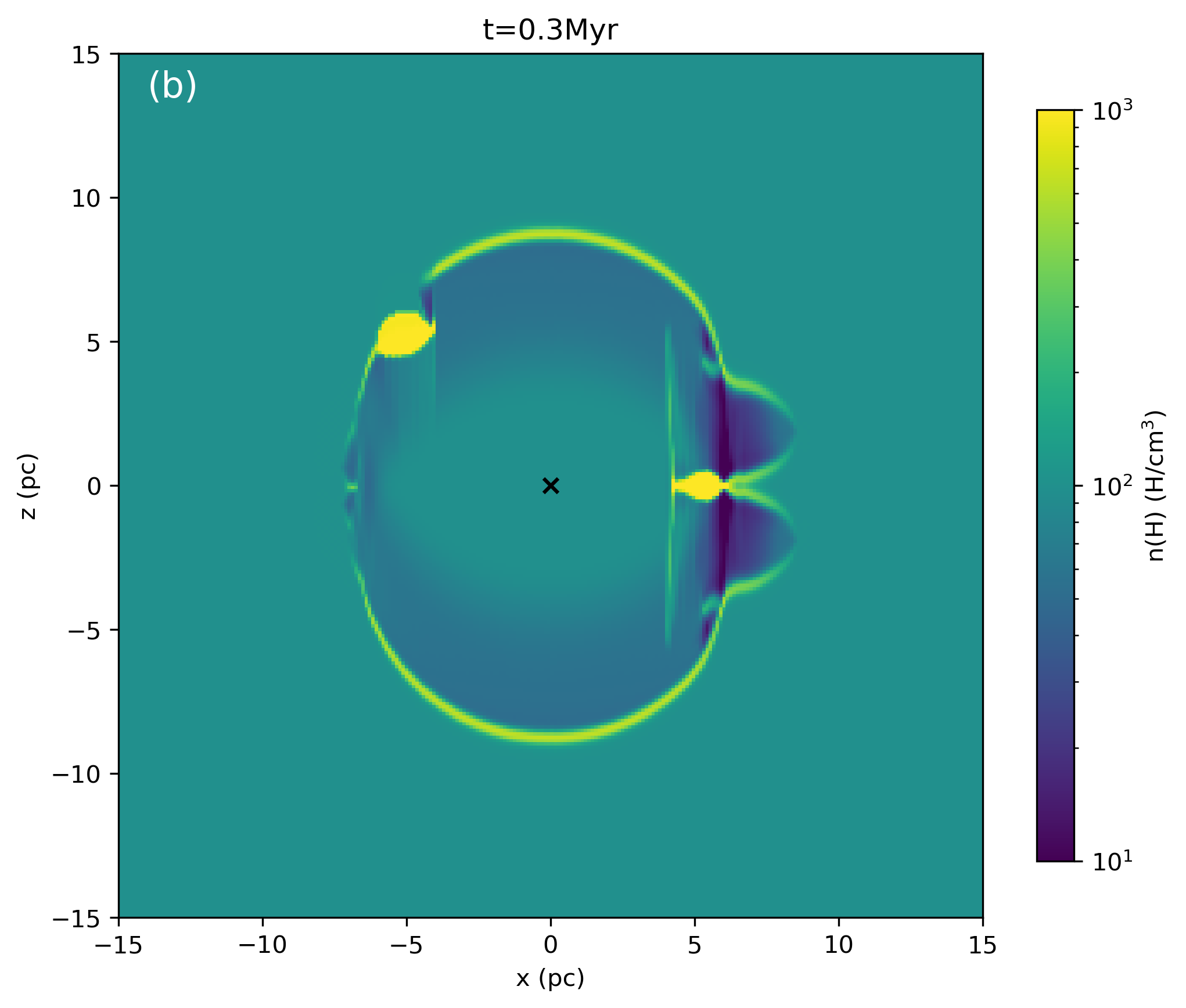}
        \includegraphics[width=0.45\textwidth]{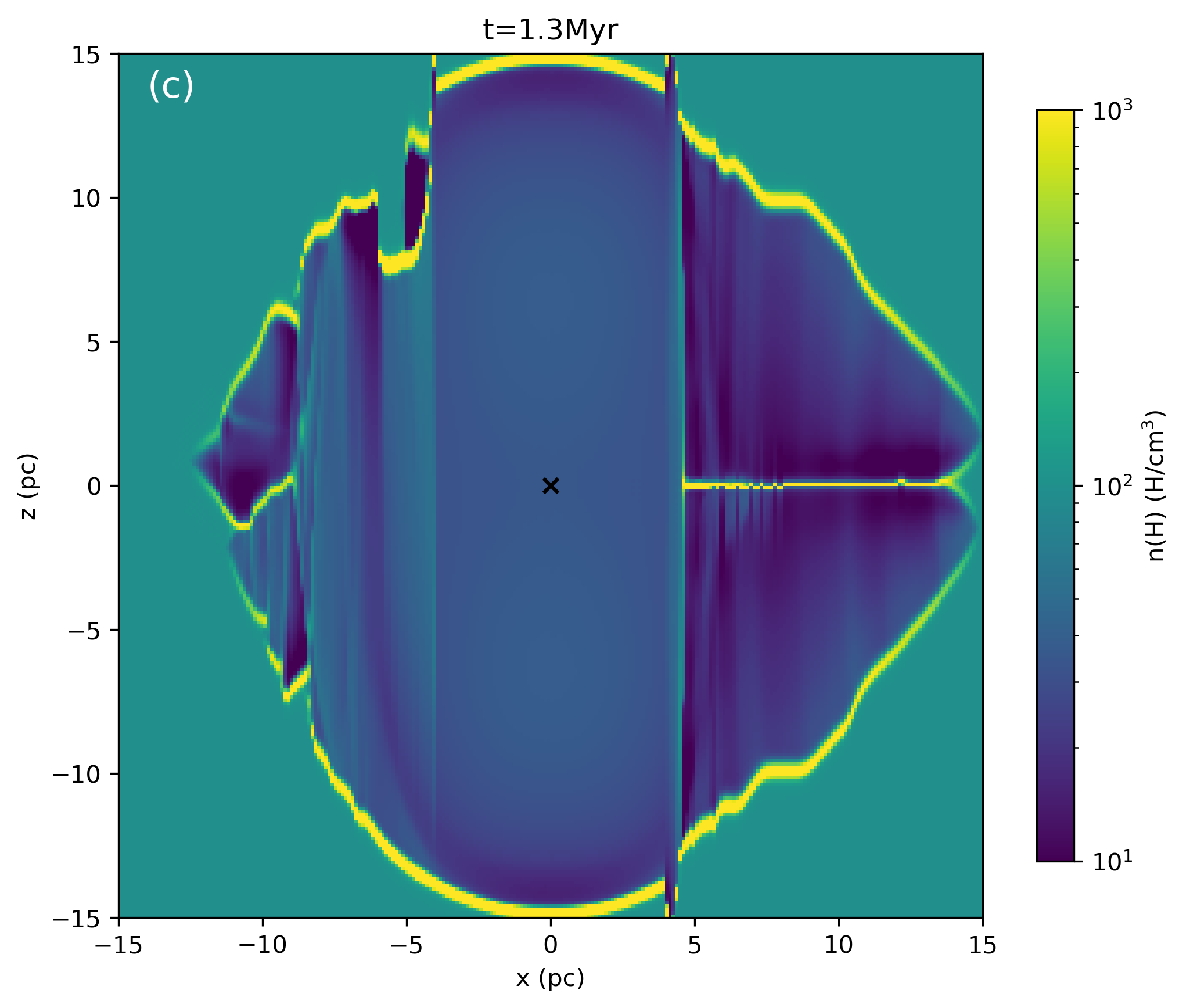}
        \includegraphics[width=0.45\textwidth]{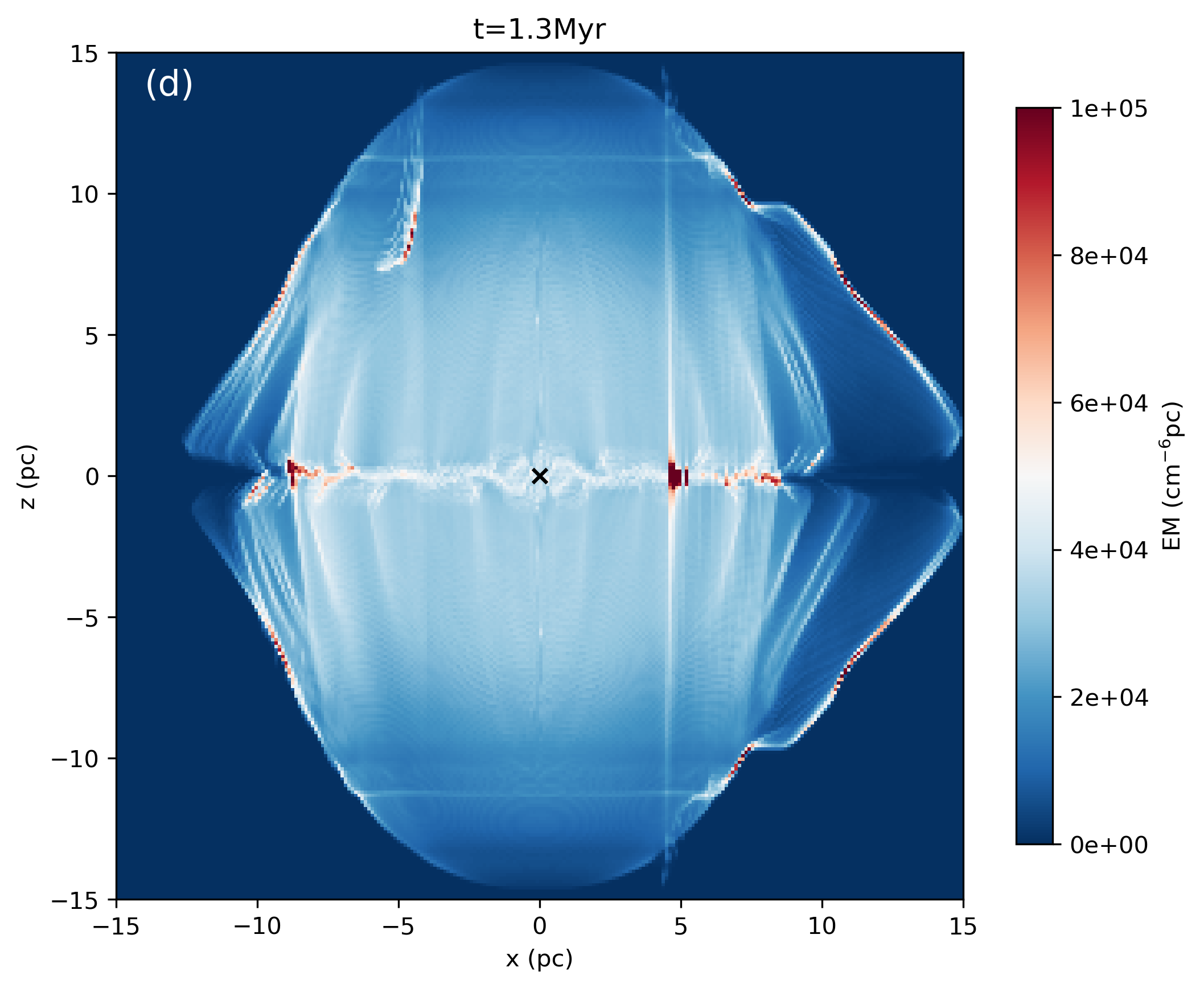}
        \includegraphics[width=0.45\textwidth]{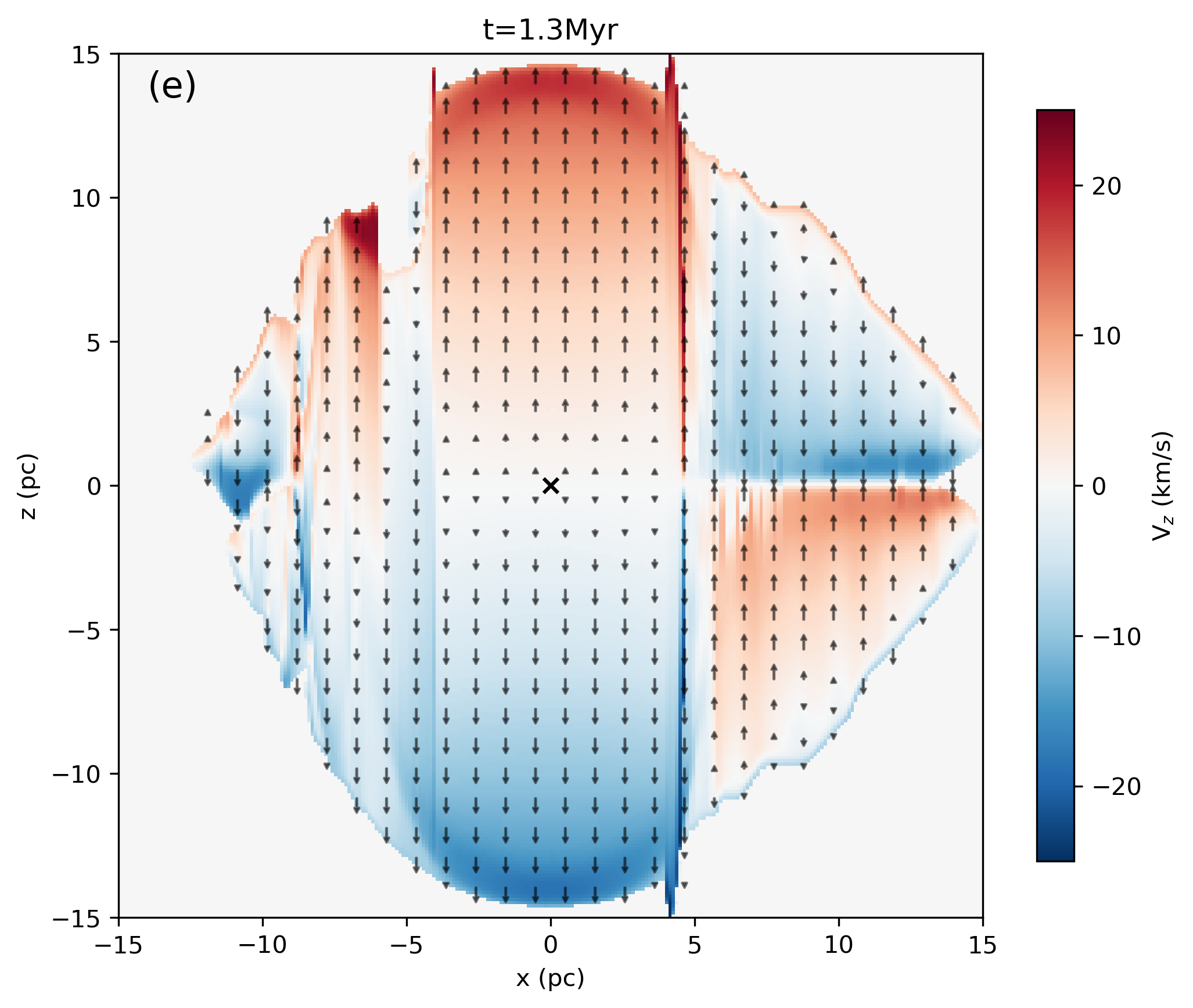}
        \includegraphics[width=0.45\textwidth]{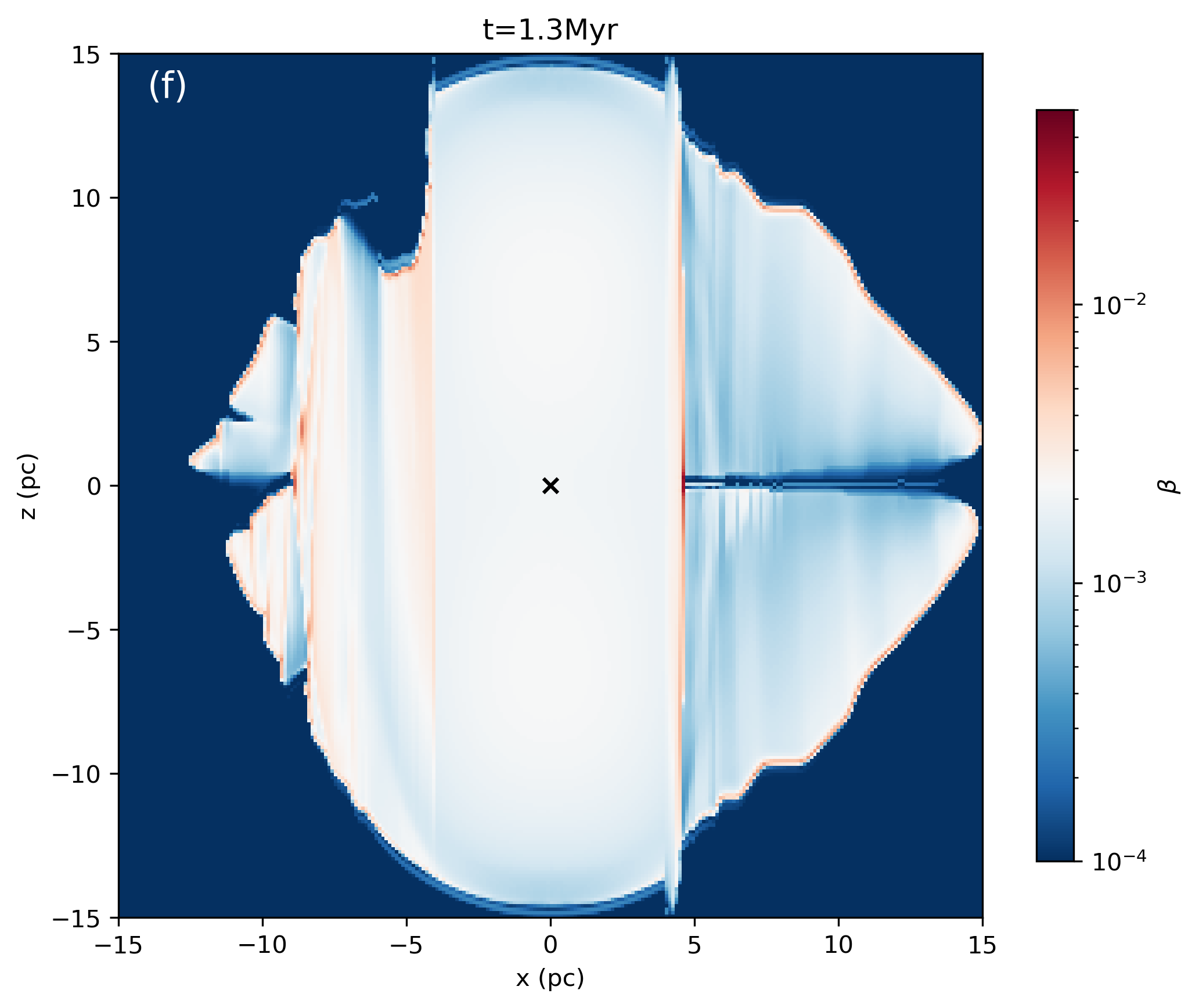}
        \caption{\label{fig:blob_slices} Select cutaways from the simulation run with preset blobs. Panel (a) shows an $x-z$ slice (along $y=0$) of the number density at $t=0$, i.e. the initial condition before activation of the ionizing source. Panels (b) and (c) show the same at $t=0.3$ and $t=1.3$ Myr, respectively. Panel (d) shows the emission measure of ionized gas along the line-of-sight. Panel (e) shows the z-velocity, $v_z$, along an $x-z$ slice ($y=0$) with velocity vectors shown in black and scaled to the magnitude of the velocity. Panel (f) shows an $x-z$ slice (along $y=0$) of the plasma $\beta$, or thermal pressure divided by magnetic pressure. The location of the ionizing source is marked with a black cross in all panels. %{\color{red} [How about making it 3 horizontal by 2 vertical, so it does not overrun the space and the upper three are all number densities?]}
        }
        
    \end{figure*}
    
    Figure \ref{fig:blob_slices} shows various quantities of the simulation, along the $x-z$ plane, in the first model, with pre-set blobs. The first three panels, (a), (b), and (c), show cuts along the $y=0$ plane, where the blobs are placed, of the number density, $n(\mathrm{H})$, at three different times. At 0.3 Myr (b), two effects are noticeable in the blob at $z=0$. Firstly, the blob has begun to be compressed along the $z=0$ plane; by 1.3 Myr (c), the blob has been flattened out into a one-dimensional stream of gas along $z=0$. Secondly, this stream is comprised of largely neutral gas.
    %{\color{red} [this can misleading: the blob is indeed flattened, but the flattened blob is only a small part of the one-D stream at $z=0$. The remaining concentration at $z=0$ is the concentration of ambient gas. One role of the blob is to create a shadow so the ``zip'' effect can happen.]}.
    Panel (d), which shows the emission measure of \textit{ionized} gas along the line-of-sight, defined as
    \begin{equation}\label{eq:emission_measure}
        \mathrm{EM=}\int n_e^2dl
    \end{equation}
    demonstrates this point. Much of this stream is likely made up of neutral gas from the ambient medium that is shielded from ionizing radiation by the blob, rather than gas from the blob itself. Secondly, by 0.3 Myr, a bright, filamentary stream of dense ionized gas has formed at the tip of the blob, at $x\approx4$ pc, about 10 pc in length. By 1.3 Myr, this filament of ionized gas has almost expanded to the length of the simulation domain ($\approx 25$ pc long, see panels [c] and [d]). However, the densest (and therefore brightest) part of the filament is heavily concentrated in the innermost parsec, best seen in panel (d).

    Panel (e) shows the z-velocity, ${v_z}$, %{\color{red} [$v$ is lowercase in equ. 1 through 4, so perhaps use $v_z$. Please make sure the notation is consistent throughout the paper]}
    with arrows showing the velocity vectors. 
    %{\color{blue}[The line for each arrow can be much shorter and the number of arrows can be greatly increased to show the velocity field better. I can show what Chun-Yen did in his paper.]} 
    A velocity enhancement is apparent at $x\approx4$ pc, the location of the filament: above $z=0$, the velocity is large and positive (around 20 km~s$^{-1}$) and below $z=0$, the velocity is large and negative (around -20 km~s$^{-1}$). This implies that the filamentary structure is in fact a \textit{flow} of ionized gas, rather than a static accumulation; in fact, the flow is supersonic, with mach numbers ranging from $1.5\sim3$ along the length of the filament. The lower right panel shows the plasma $\beta$ parameter,
    \begin{equation}\label{eq:plasma_bbeta}
    \beta=P_T/P_B
    \end{equation}
    where $P_T$ and $P_B$ are the thermal and magnetic pressures, respectively. $\beta<1$ across the entire simulation domain, implying that all of the ionized plasma is magnetically dominated. Within the ionized gas filament, $\beta$ is enhanced, ranging from a few hundredths at the ``tip'' of the filament to $\sim0.005$ across the rest of its extent. %{\color{blue}[I am surprised that beta is much lower than $10^{-3}$ (the initial value) on the equator, given that region is compressed and should be denser than the initial ambient density. Please double check. Also, should we comment on why the placement of the blobs is important? The z=0 blob is confined and compressed by the photo-evaporative flows from above and below the midplane and is enhanced by the mirror symmetry, which may not be there in real clouds. The lack of symmetry allowed the off-middle-plane blob to be pushed away together with the ambient medium, forming a corrugated surface that facilitated the "sheet" formation.]}

    %{\color{blue}[Comments on Fig.1: Left bottom: use one color for positive and another for negative vz, as in Chun-Yen's new paper? I will show it on Friday. Right bottom: Reduce the color range to make the pressure gradient more obvious. ]}

    These last two observations, the low plasma $\beta$ of the filament and its high velocity, outflowing from the tip of the neutral gas blob, implies that the filament is a \textit{magnetically confined plasma flow}, a flow of ionized gas that has been stripped off (photoablated) from the surface of the dense neutral gas blob and forced to flow in line with the magnetic field in the $\hat{z}$-direction. The enhancement of $\beta$, and therefore the thermal pressure, at the tip of the blob (panel [f]), is also noteworthy, implying a negative pressure gradient out to increasing values of $|z|$ that causes the filament to expand along the direction of the magnetic field.

    Interestingly, the left blob in the simulation, which is identical to the right blob except for its placement at $(x,y,z)=(-5,0,5)$, displays a markedly different morphology, particularly at later times. At $t=0.3$ Myr (panel [b]), a slight density enhancement in the ionized gas can be seen on the right side of the blob as it is being photoionized, but by 1.3 Myr (panels [c], [d]), it becomes clear that the flow of plasma from this blob resembles a thick, two-dimensional ``sheet'' (or three-dimensional ``cylinder'') rather than a one-dimensional ``filament''. %{\color{red} [Is it possible that the 3D structure is a cylinder?]} 
    This is due to the fact that the entire bottom face of the blob is exposed to the ionizing source, meaning that the flow of ionized gas is much more extended along the $x$-axis than the flow of plasma from the blob at $z=0$. The slight enhancement of thermal pressure at the tip of the blob (panel [f]) suggests that this flow of ionized plasma is indeed caused by photoionization of the outer layer of the blob, similar to the blob at $z=0$. Clearly, the formation of a filament, in this simulation setup, depends closely on the geometry of the configuration. The $z=0$ blob is confined and compressed by the photo-evaporative flows from above and below the midplane, and the resulting filament formation is enhanced by the mirror symmetry along the z-axis, which may not be present in real clouds. Lack of symmetry about the z-axis caused the off-middle-plane blob to be pushed away together with the ambient medium, forming a corrugated surface that facilitated the formation of a ``sheet'' rather than a ``filament''. However, it is noteworthy that a curved filament-like formation is still visible in the emission measure map (panel [d]) at the ionization interface of the blob.
    
    \subsection{Simulation with driven turbulence}\label{sec:turbulence_simul}
    \begin{figure*}
        \includegraphics[width=0.49\textwidth]{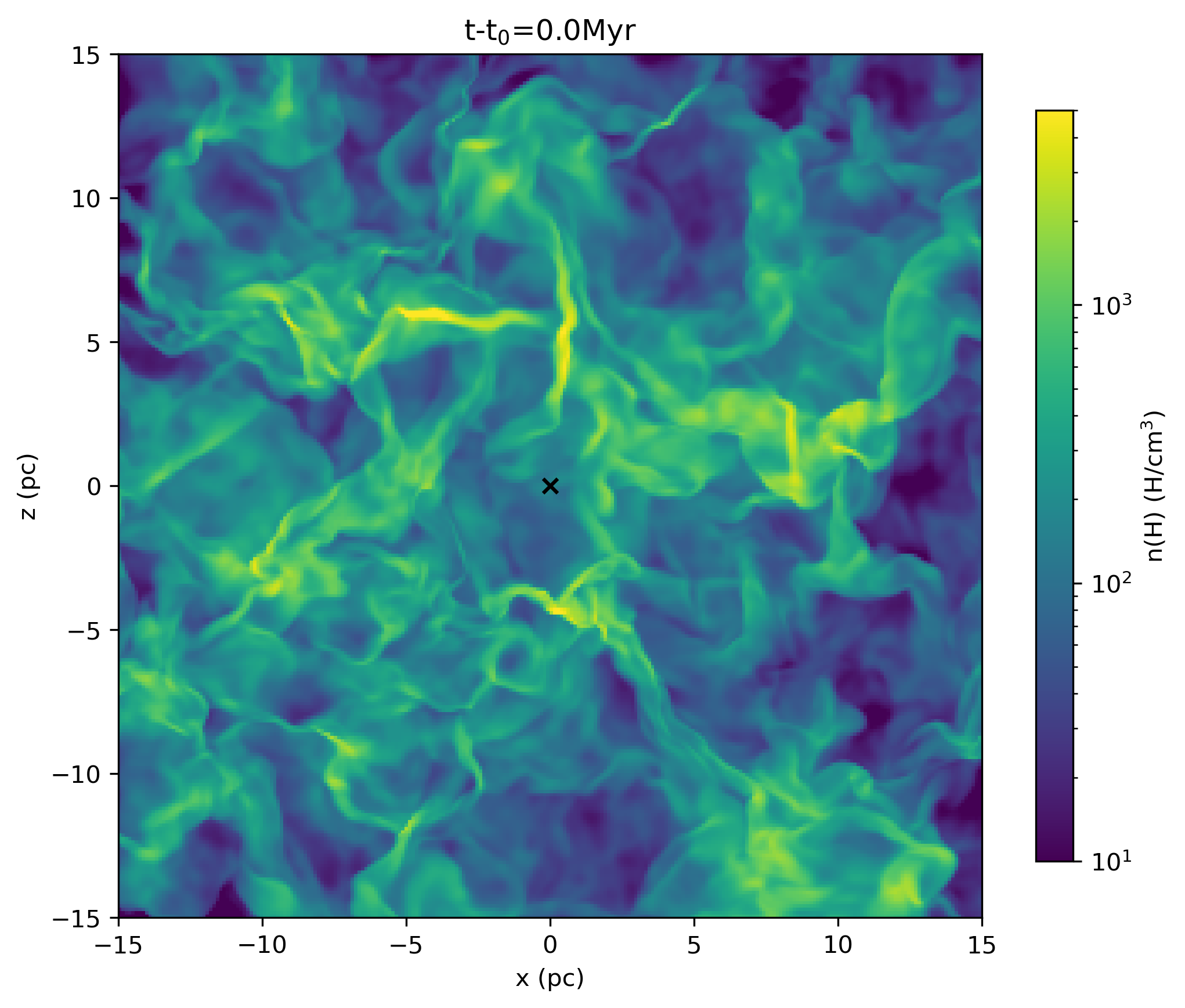} 
        \includegraphics[width=0.49\textwidth]{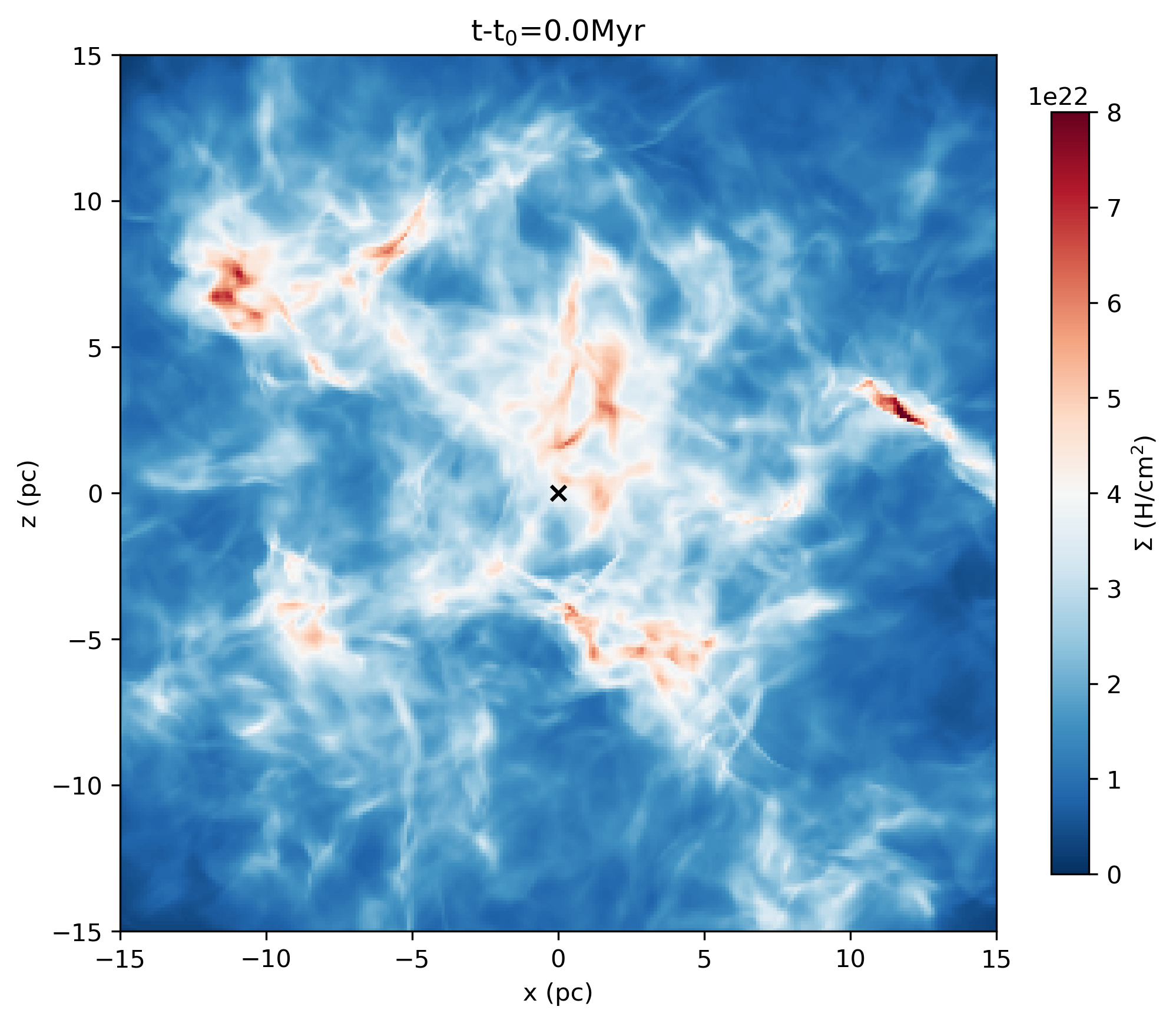}
        
        \caption{\label{fig:turb_t=0_slices} Gas density along the $x-z$ plane at the time of activation of the magnetic field and ionizing source ($\mathrm{t_0}$) in the simulation run with driven turbulence. The left panel shows a slice of the gas density in the $y=0$ plane, and the right panel shows the integrated gas density along the line-of-sight. The ionizing source is marked with a black cross in both panels.}
    \end{figure*}
    We now examine the second simulation run, where a period of driven turbulence has been used to seed an initial condition with a strong density contrast. Figure \ref{fig:turb_t=0_slices} shows two panels, along the $x-z$ plane, at the time the magnetic field and ionizing source are activated and the driving of turbulence ends (denoted as $\mathrm{t}_0$, or 20 Myr). The left panel shows the total number density, $n$(H), in the $y=0$ plane, and the right panel shows the column density of gas along the line-of-sight. As the ionizing source has just been activated, all gas is neutral at this stage. Note the large fluctuations (from $<10$ to a few $\times10^3$ H/cm$^{3}$) in the ambient medium compared to the first simulation setup, which assumed a completely uniform initial density (except for the two pre-set spherical blobs). This creates pockets of low- and high-density that %emulate 
    mirror the pre-set blobs considered previously.%, but in a more organic environment.

\begin{figure*}
        \centering
        \includegraphics[width=0.45\textwidth]{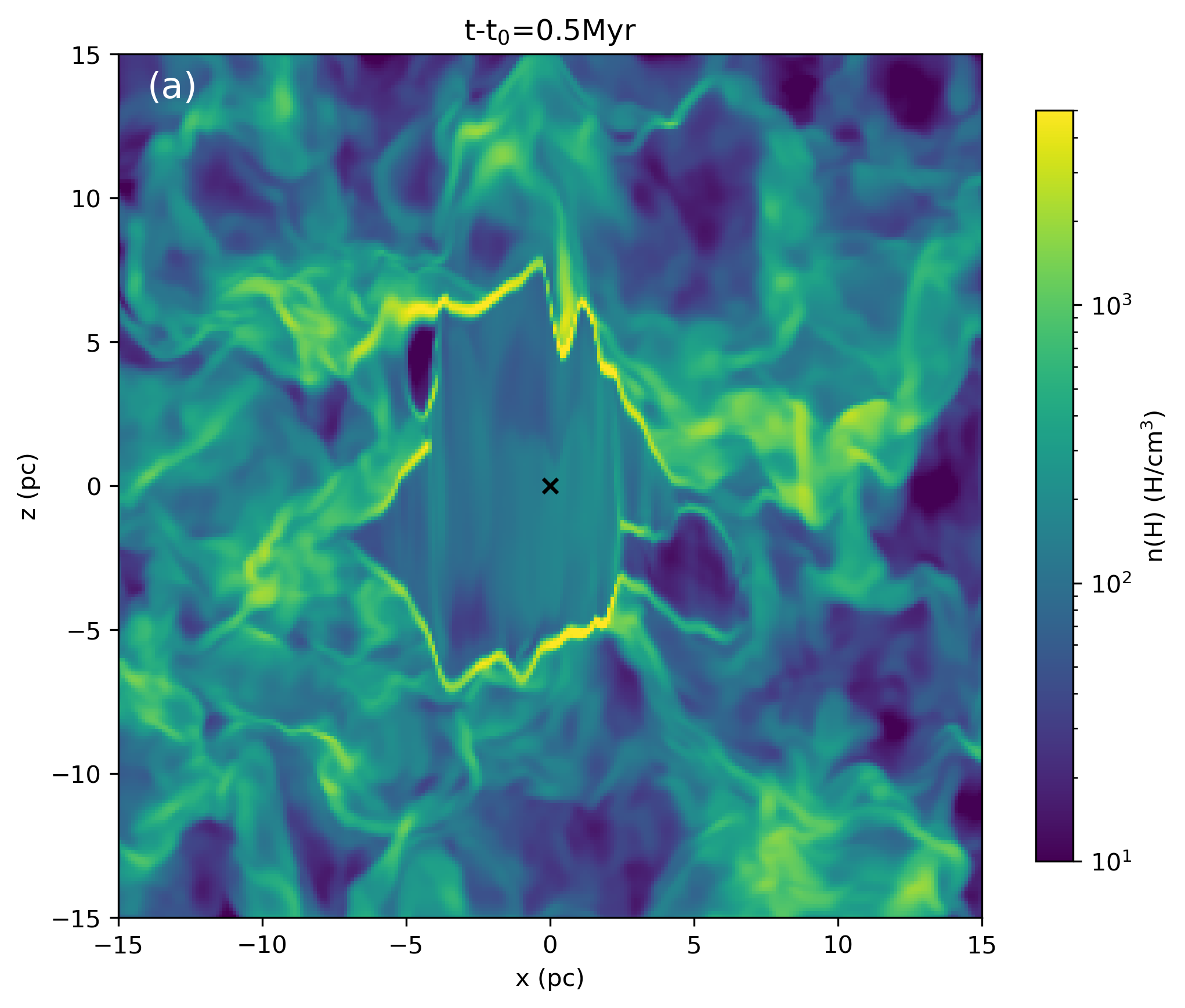}
        \includegraphics[width=0.45\textwidth]{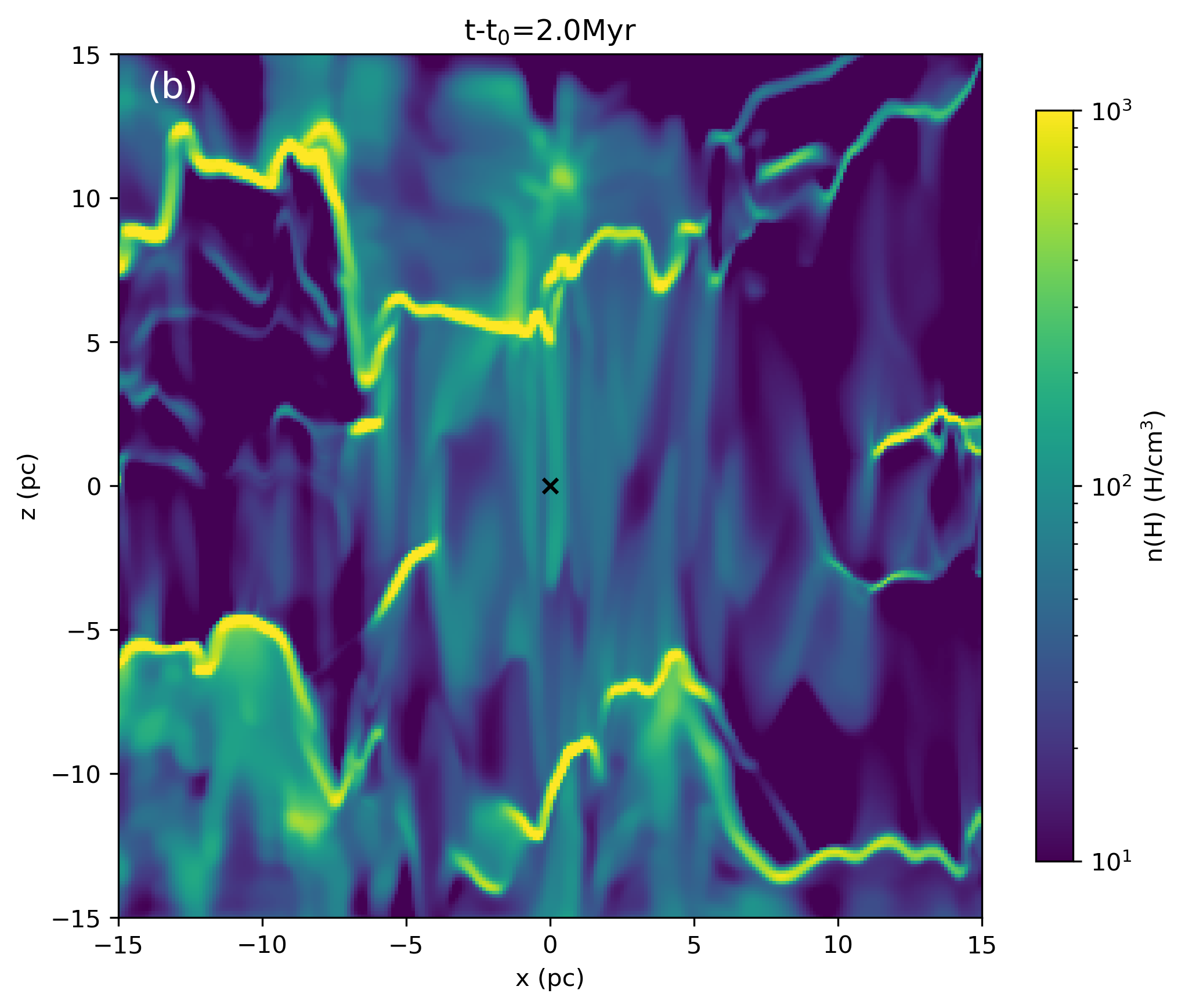}
        \includegraphics[width=0.45\textwidth]{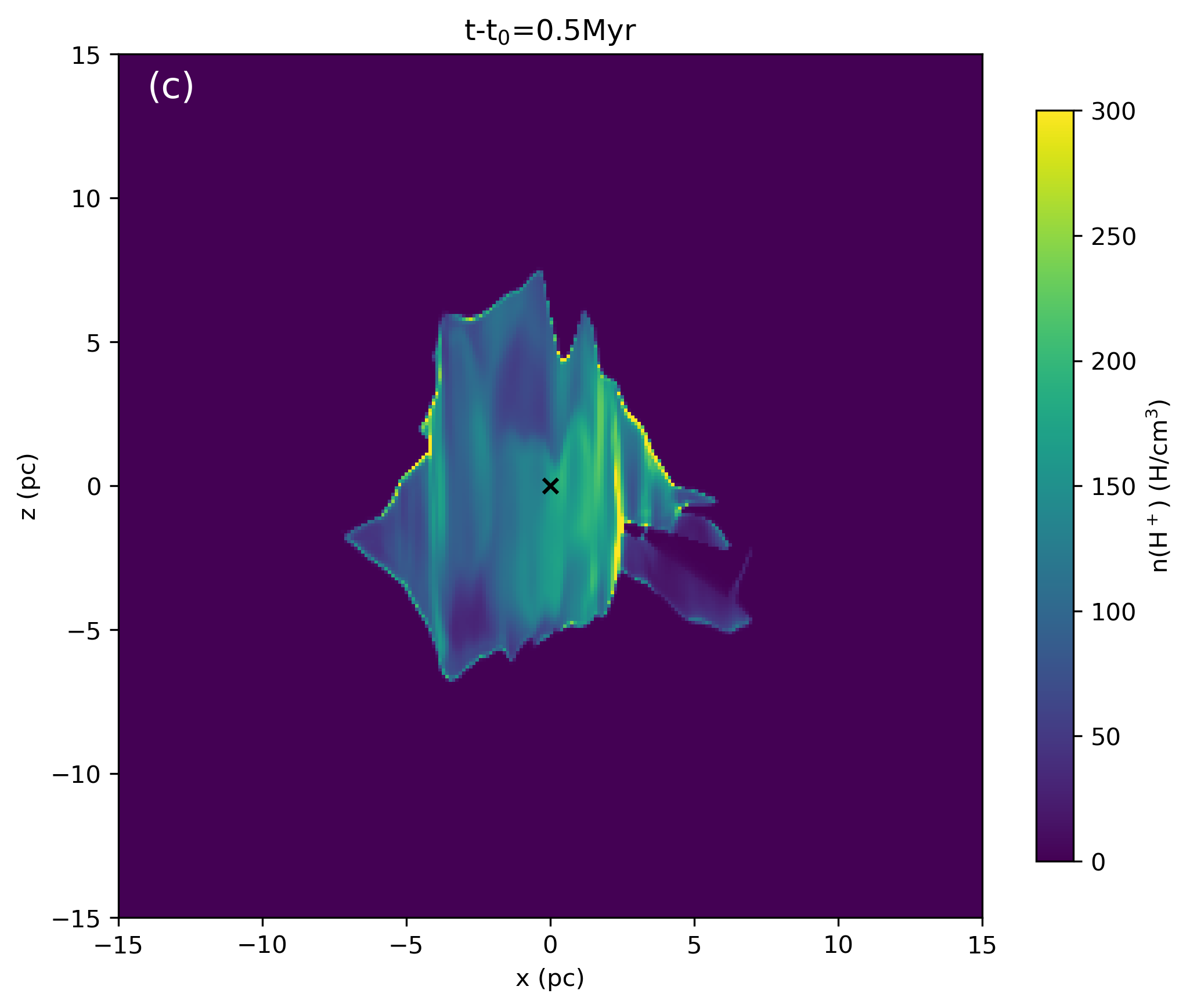}
        \includegraphics[width=0.45\textwidth]{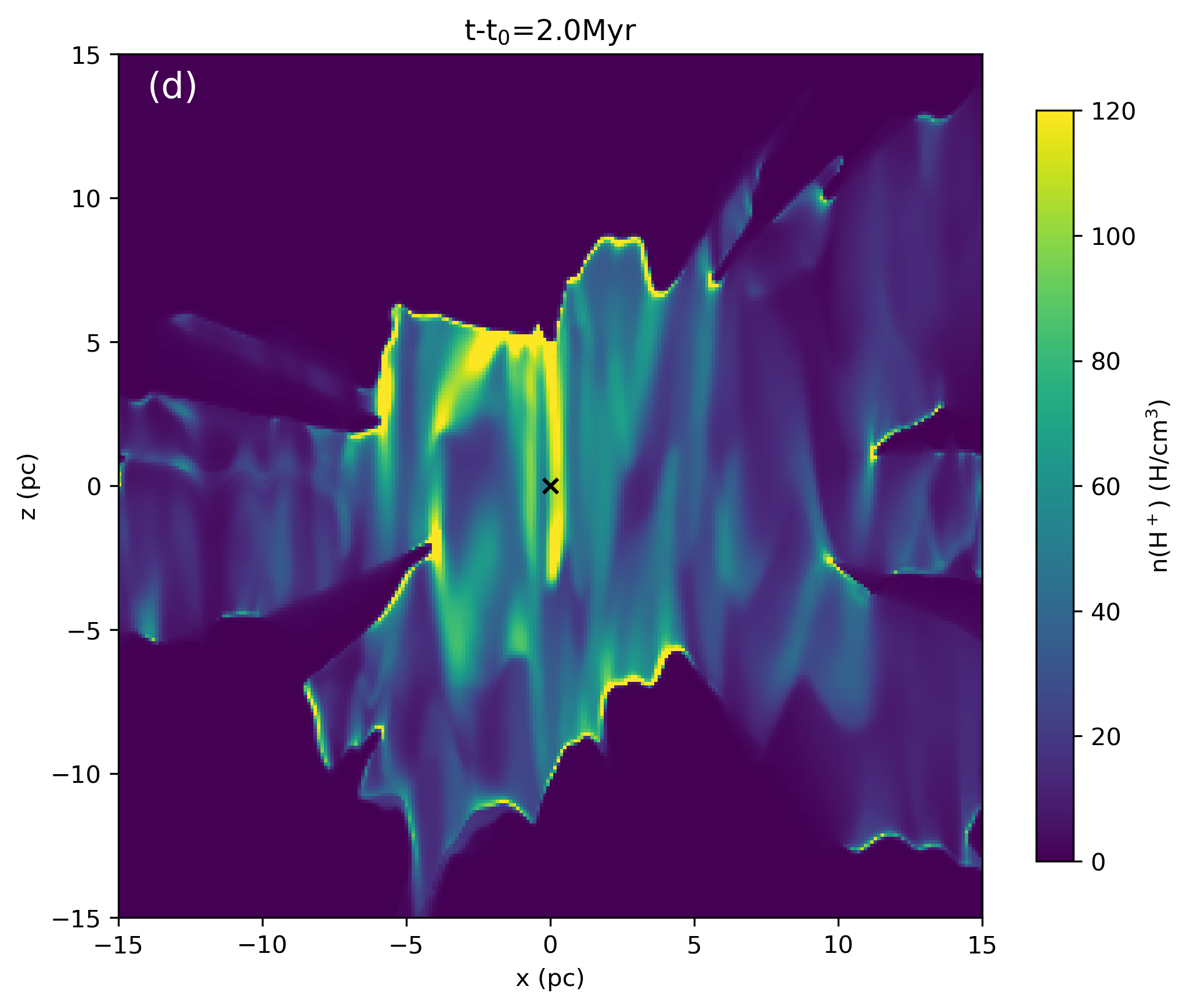}
        \includegraphics[width=0.45\textwidth]{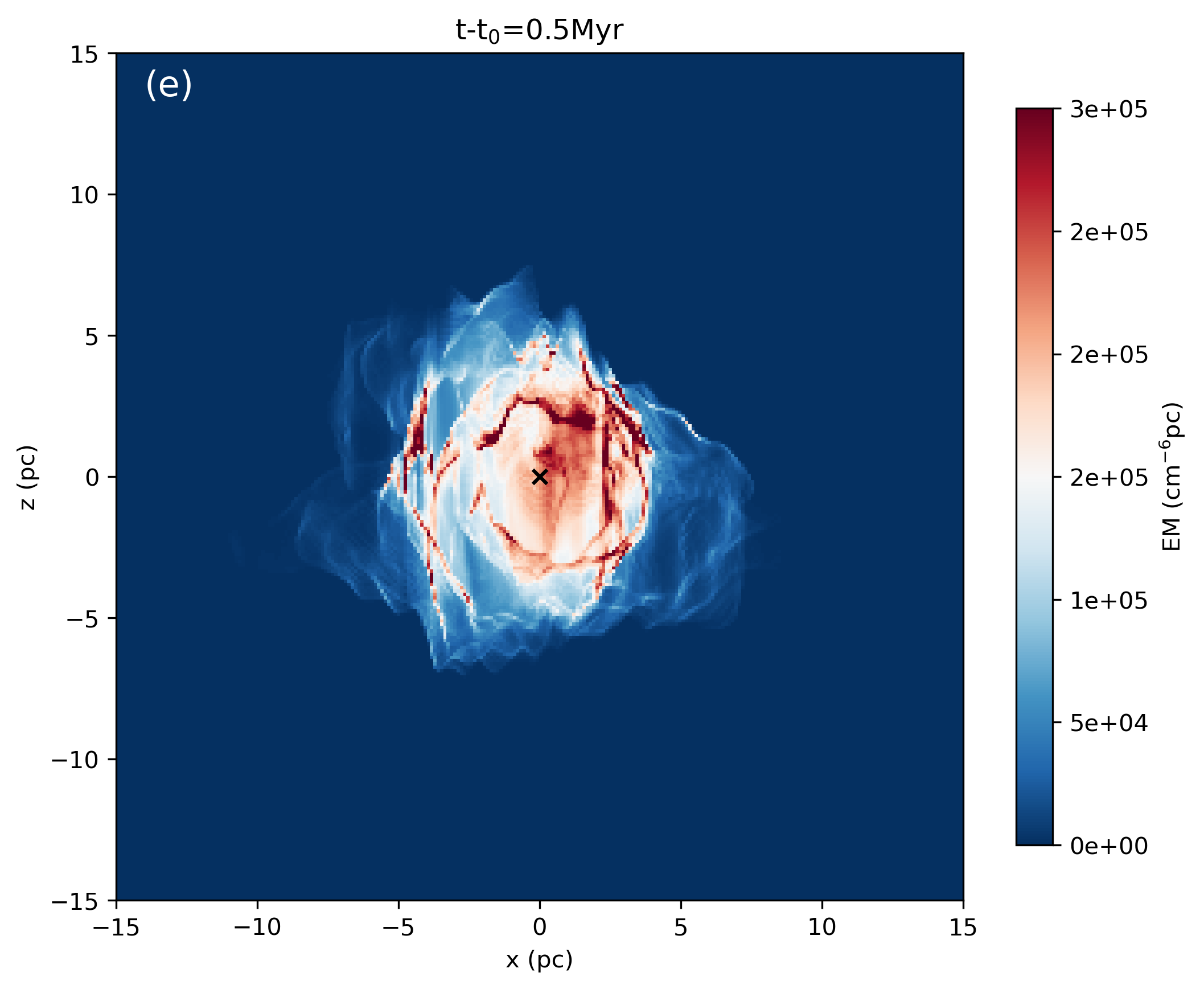}
         \includegraphics[width=0.45\textwidth]{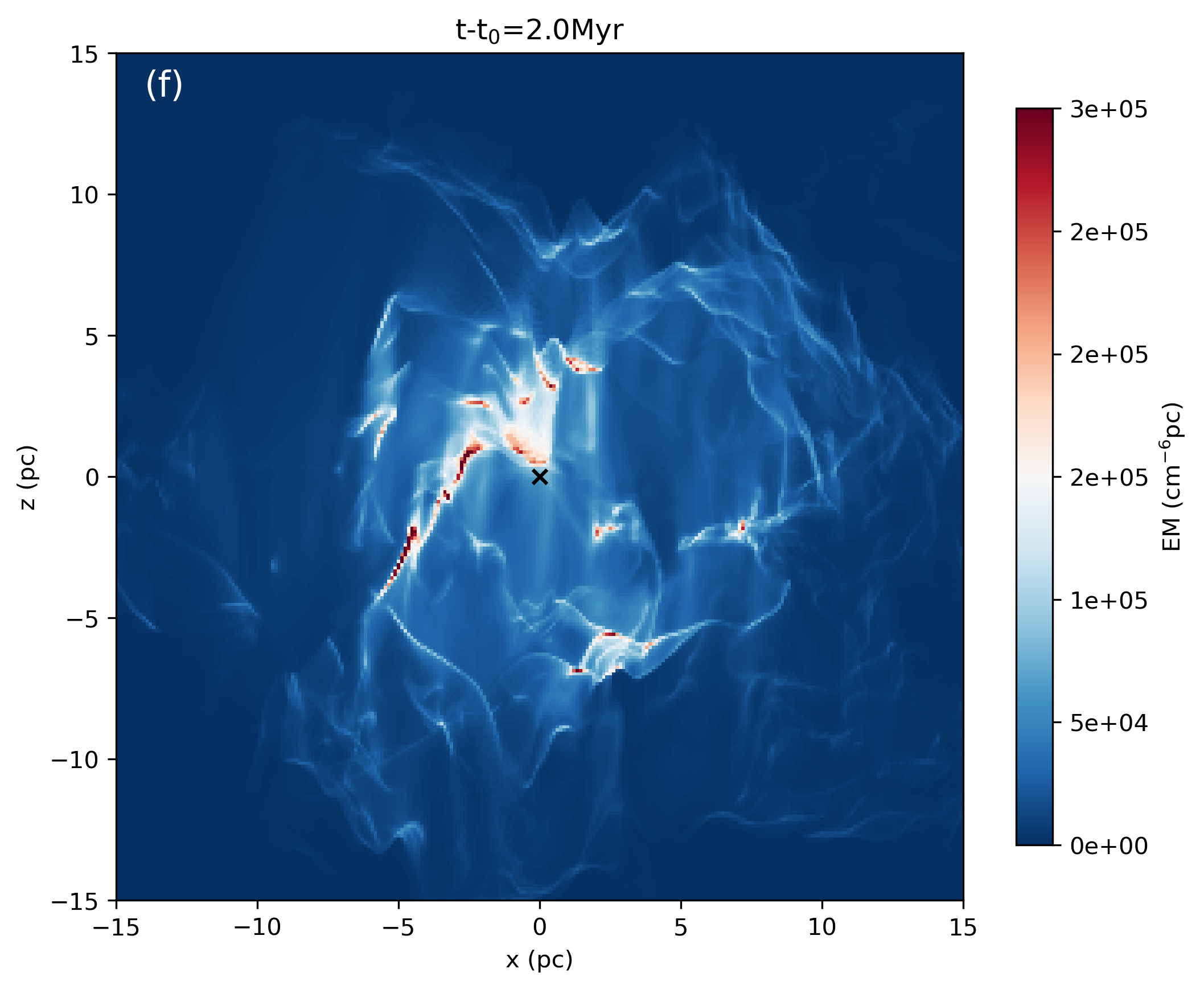}\caption{\label{fig:turb_t!=0_slices} Select cutaways from the simulation run with driven turbulence. Panels (a) and (b) show $x-z$ slices (along $y=0$) of the \textit{total} number density of gas at $\mathrm{t-t_0}$=0.5 and 2 Myr, respectively. Panels (c) and (d) show the same but for the \textit{ionized} gas number density. Panels (e) and (f) show the emission measure of ionized gas along the line of sight. The location of the ionizing source is marked with a black cross in all panels.}
    \end{figure*}

    Figure \ref{fig:turb_t!=0_slices} shows several panels, along the $x-z$ plane, of the density of ionized and neutral gas at two different times after activation of the ionizing source: 0.5 Myr and 2 Myr. Panels (a) and (b) show the \textit{total} number density of gas (i.e., neutral plus ionized) along the $y=0$ plane, whereas panels (c) and (d) show the number density of \textit{ionized} gas along the $y=0$ plane. Panels (e) and (f) show the emission measure of ionized gas along the line-of-sight.

    At $\mathrm{t-t_0}$=0.5 Myr, ionized gas filamentation is apparent in two distinct configurations (see panel [c]). Along $x\approx2.5$ pc, a dense filament extends along the z-axis, originating from the tip of a dense ``pillar'' of neutral gas that extends into the \Hii\:region. This ionized gas filament seems to have been generated by a similar mechanism as the filament located at $x\approx4$ pc shown in Fig. \ref{fig:blob_slices} and discussed in \S\ref{sec:blob_simul}, with the dense ``pillar'' playing a nearly identical role as the dense blob placed at $z=0$ in the first simulation. This filament is also apparent in the emission measure map (panel [e]) and shows the same velocity signature as shown in Fig. \ref{fig:blob_slices} panel (e), but with mach numbers closer to $\approx1$ across the filament. %{\color{red}[If the pillars are trans- or super-sonic, would you expect to observe shock in the filaments?]}
    
    Other, lower density filaments are apparent as well, including the filament (or pair of filaments) at $x\approx-4$ and $x\approx2$ pc. These filaments do not have an obvious pillar or blob to feed them, but rather are generated by specific geometries in which a ``wall'' of dense neutral gas at constant $x$ is photoionized (e.g., the wall extending from $x\approx-4$, $z\approx3\sim6$ pc, and from $x\approx2$, $z\approx4\sim5$ pc in panel [c]), causing an overdensity of magnetically-confined plasma to flow along a filament at constant $x$. Despite resulting from different configurations, in underlying mechanism these two types of filaments are essentially identical: both are magnetically-confined plasma flows.

    At $\mathrm{t-t_0}$=2 Myr (panels [b], [d], [f]), the \Hii\:region has evolved and expanded to the extent that several more pillars (and filaments) are apparent, including those at $(x,z)\approx(-5,-2)$, $(x,z)\approx(-6,3)$, $(x,z)\approx(6,7)$, $(x,z)\approx(10,-2)$, $(x,z)\approx(11, 1)$, etc. The emission measure map, panel (f), shows a highly filamentary morphology, including filaments around the pillars mentioned above.

    Many of the filaments at this time step are curved, rather than almost perfectly linear like the filament in Fig. \ref{fig:blob_slices}. This can be interpreted as the result of rapid photoionization of the pillars; since essentially all of the plasma seeding a given filament comes from the tip of the pillar, as the tip retreats away from the ionizing source because of continuous photo-ablation, it moves across magnetic field lines, making the source of the plasma flow in the ionized filament shift across the field lines as well. For a specific illustration and explanation of this phenomenon, see Appendix \ref{sec:curved_fils}.%{\color{blue}[Would an illustration with a blob be worthwhile in the appendix?]}

\section{Discussion}\label{sec:discussion}

    \subsection{Comparison with Sgr C}\label{sec:sgr_c_comparison}
        As discussed in \S\ref{sec:introduction}, this study was motivated by JWST observations of ionized gas filamentation in %the Sgr C star-forming region
        a highly-magentized star-forming region in the Central Molecular Zone, Sgr C %which are detailed by 
        \citep{bally25}. It must first be noted that the physical setup in the simulation is somewhat artificial and therefore not entirely representative of a real star-forming region like Sgr C. In particular, the fact that the cloud was evolved \textit{without} magnetic fields, only under the effect of driven turbulence, before the input of the ionizing source and a strong (purely vertical) magnetic field, likely does not reflect the formation history of an actual CMZ cloud. Therefore, the use of our simulation to compare with Sgr C will largely be limited to the effect of strong magnetic fields on gas morphology. %with the simulation results presented in \S\ref{sec:results} 
        %is warranted. 
        %We choose to concern ourselves with the second simulation run, which includes a period of driven turbulence, as we view this simulation as closer to a real environment.

        The most direct comparisons can be made with the length and emission measure of the filaments in Sgr C. We choose to compare the emission measure rather than the density because of the inability to probe the line-of-sight depth of features in Sgr C and the vast differences in spatial resolution between the simulations and observations ($\approx0.006$ pc with the JWST, 0.12 pc with our simulations). With respect to the former point, we note that the filaments in our simulations are roughly cylindrical (rather than sheet-like), supporting the cylindrical approximation made in Sgr C \citep{bally25}. The latter point is more intransigent, and suggests that any density derived from emission measure in the simulation would be a severe underestimate because of the overestimation of the line-of-sight depth (since $n_e\propto L^{-0.5}$).

        The ionized gas filaments in Sgr C have background-subtracted emission measures ranging from a few $\times10^4$ to $\approx5\times10^5\:\mathrm{cm^{-6}pc}$, depending on the adopted extinction value, and are up to around a parsec in length \citep[see Table 2 of][]{bally25}. The filaments in our simulations show similar values. In the first simulation, with pre-placed blobs, the brightest part of the filament at $x\approx4$ (see Fig. \ref{fig:blob_slices} [d]) has a background-subtracted emission measure of $\approx6\times10^5\: \mathrm{cm^{-6}pc}$ and is around a parsec in length, similar to the filaments in Sgr C. It is worth noting that the rest of this filament, which extends for almost the entire simulation domain, has a significantly lower emission measure, $\approx2\times10^4\:\mathrm{cm^{-6}pc}$, which is below the lower end of emission measure values in Sgr C. This implies that only the innermost parsec of this filament would be visible in Sgr C. Inversely, this result suggests that the filaments in Sgr C may in fact be much longer than were observed simply due to the sensitivity limits of the JWST observations. The filaments in the simulation run with driven turbulence also show a similar emission measure to the filaments in Sgr C.

        The shape, extent, and age of the simulated \Hii\:region and Sgr C are also worth comparing. The ``core'' of the brightest emission in the Sgr C \Hii\:region measures around 7 pc on its long dimension (roughly parallel to the galactic plane) and around 5 pc on its short dimension (perpendicular to the galactic plane). The ``core'' of the most dense portion of the simulated \Hii\:region is around 10 pc in all dimensions, or about twice as large as Sgr C, although there is some slight preferential elongation along the $x-$ and $y-$axes at later times (see panels [d], [f] of Fig. \ref{fig:turb_t!=0_slices}). The $\approx2\times$ difference in overall size may be attributable to the $\approx2\times$ higher average density in Sgr C compared to the simulated medium, though many other factors may be at play. %The slightly lower density for the simulation runs was chosen as a practical consideration, as a more dense, compact \Hii\:region would be more computationally intensive to model (particularly in the demands on increased resolution) compared to a diffuse, extended one. Future simulations may be able to more accurately model in high-resolution an environment with a Sgr C-like density.

        There are some striking similarities in the shape of the two \Hii\:regions. As mentioned, both the simulated and observed \Hii\:regions are nearly spherical (circular in the case of Sgr C), but slightly elongated \textit{perpendicular} to the magnetic field lines \citep[if, in the case of Sgr C, the poloidal component of the CMZ magnetic field is the dominant one, which is tenuously suggested by the statistical preference of filament orientations along the poloidal component of the field; see Fig. 6 of][]{bally25}. This result is quite unexpected given the suggestion of several previous studies that simulated magnetized \Hii\:regions are distorted and elongated \textit{parallel} to the field lines \citep{krumholz07,arthur11,zamoraaviles19,gendelev12}. It is possible that this phenomenon, the elongation of the \Hii\:region perpendicular to the field, could be due to the lifting of material away from the midplane by the strong perpendicular magnetic field, allowing the ionizing radiation to cleave through regions of higher density more easily and carve out a lower-density ``tunnel'' along the midplane (see panels [b], [d] of Fig. \ref{fig:turb_t!=0_slices}). However, this is only one possible mechanism. %{\color{blue} [Let us discuss this effect on Friday; it is indeed unexpected and I am not I understand why.]}

        Another interesting feature is the presence of a diffuse, filamentary ``halo'' of ionized gas around the ``core'' of the \Hii\:region, which is seen in both Sgr C and the simulation. This is best seen in panel (f) of Fig. \ref{fig:turb_t!=0_slices} and the top panel of Fig. 4 of \citet{bally25}. This diffuse halo implies leakage of ionizing photons into the extended medium surrounding each \Hii\:region; in the case of the simulation with driven turbulence, this is likely due to the ease with which the ionization front eats through regions of low density, forming the pockets of ionized gas best seen in panel (d) of Fig. \ref{fig:turb_t!=0_slices}.

        Finally, one interesting comparison between the filaments in Sgr C and the simulated \Hii\:region is the morphology of the filaments themselves. The filaments in Sgr C are generally more numerous and have a wider diversity of curvature and orientation than the filaments in the simulated \Hii\:region. Both differences (particularly in number of filaments) are likely attributable in large part to discrepancies in resolution between the observations and simulation. Differences in curvature are likely due to lower density in the blob/pillar/wall that feeds the filament, causing the source of the plasma flow to shift as the reservoir of neutral gas is rapidly eroded away; this is explored further in Appendix \ref{sec:curved_fils}. Differences in orientation are harder to treat, as they imply complicated line-of-sight effects in Sgr C that are difficult to model with the constraints on resolution and magnetic field orientation in the simulation setup used in this work. Particularly, the filamentation in Sgr C often displays a pattern of overlapping or nearly-overlapping filaments with different orientations, at times even in a cross-hatching pattern \citep[see Fig. 5 and Appendix A of][]{bally25}. This implies %rapidly 
        locally varying magnetic fields, both across the plane of the sky and along the line-of-sight, that have been suggested in the literature \citep[see, e.g.,][]{Tress2024} and are not modeled in the simulations presented in this paper. Future high-resolution mapping of the magnetic field in magnetically-dominated \Hii\:regions, like Sgr C, may further illuminate the effect on local magnetic field variation on filament formation.
        
        Also important are measurements of the velocity of the ionized gas, since the model for filament formation given in the present study rests on the \textit{motion} of the ionized gas, particularly at speeds that meet or exceed the sound speed. Therefore, high spatial- and velocity-resolution observations of the ionized gas in Sgr C, and other CMZ \Hii\:regions, will be crucial to testing our model. Acceleration of the ionized gas may be an important consequence of the incorporation of magnetic fields into \Hii\:region models, and contribute to the surprisingly high velocities observed in some \Hii\:region complexes \citep[e.g., Orion,][]{pabst19}.

        %\subsection{Implications of ionized gas filamentation}\label{sec:filaments_implications}

        %\textbf{Not quite sure what to put here yet... I am imagining some discussion of the potential broader implications of ionized gas filamentation on stellar feedback, galactic gas dynamics, escape fraction, shocks in clouds, particle acceleration, etc. But I have not condensed my thoughts enough (or read enough papers) to write anything here. Suggestions would be recommended.}

\section{Conclusions}\label{sec:conclusion}
In this paper, we have presented a suite of RMHD simulations of \Hii\:region expansion and evolution that effectively demonstrates the formation of ionized gas filaments
%, similar to those observed by \citet{bally25}, 
in strong magnetic fields. We find that these filaments can be explained effectively as magnetically-confined ($\beta\ll1$) flows of photoionized plasma. These flows must be seeded by a reservoir of dense neutral gas. We present two possible geometries, one in which the reservoir is a freestanding dense ``pillar'' or ``blob'', and one in which the reservoir is a long ``wall'' of neutral gas parallel to the magnetic field. In both cases, the curvature of the filament is dependent on the density of the reservoir.

A key feature of our model for ionized gas filament formation is the motion of the ionized gas. In particular, we predict relatively high velocities, around a few dozen kilometers per second, within the filaments. Therefore, future high-spatial and -velocity measurements of ionized gas in %CMZ 
highly-magnetized \Hii\:regions will be crucial to testing our model.

%Although we are able to replicate the formation of both curved and straight ionized gas filaments, we note several discrepancies between our models and Sgr C. First, the filaments in our models are longer and less dense than the filaments in Sgr C, which is potentially due both to constraints on the simulation parameters and the environment around Sgr C. Second,  Finally, we find that, contrary to the literature, the \Hii\:region is elongated perpendicular, rather than parallel, to the magnetic field lines.

We find a favorable comparison between the length and emission measure of the simulated filaments and the filaments in Sgr C. We also observe a slight elongation of the simulated \Hii\:regions \textit{perpendicular} to the field lines that may also be present in Sgr C.
%that we are successfully able to recreate the formation of ionized gas filaments with similar length and emission measure (and therefore, likely, density) to the filaments in Sgr C in both of our simulation runs. We also successfully recreate the slight elongation of the Sgr C \Hii\:region \textit{perpendicular} to the magnetic field lines.

However, we are not able to replicate the large number and diversity of filament orientations observed in Sgr C, and the complicated structure along the line-of-sight. We expect that this will be more difficult to treat, involving greater knowledge of the local magnetic field in %Sgr C 
highly-magnetized \Hii\:regions like Sgr C and more sophisticated, higher-resolution simulations accounting for rapid variations in the local magnetic field.

\begin{acknowledgements}
\textit{Acknowledgments:} This work is based on simulations performed using computing resources from UVA research computing (RIVANNA) and NSF’s ACCESS computing allocation PHY250098. S.T.C. acknowledges support from the award JWST-GO-04147.003-A. Y.T. acknowledges support from the interdisciplinary fellowship at UVA. Z.Y.L. and Y.T. are supported in part by NASA 80NSSC20K0533, NSF AST-2307199, JWST-GO-02104.002-A, and the Virginia Institute of Theoretical Astronomy (VITA). J.-G.K acknowledges support from KIAS Individual Grant QP098701 at Korea Institute for Advanced Study.

\end{acknowledgements}

\bibliography{masterbib.bib}
\begin{appendix}
    \section{Model with lower blob densities: curved filaments}\label{sec:curved_fils}
    In order to demonstrate the mechanism by which a \textit{curved} ionized gas filament, discussed in \S\ref{sec:turbulence_simul}, could form, we present a model with identical setup to the pre-set blob simulation described in \S\ref{sec:methods}, but with blob densities of $250$ H/cm$^{-3}$ rather than $1000$ H/cm$^{-3}$, i.e. $4\times$ lower, only $2.5\times$ more dense than the ambient medium.

\begin{figure*}
        \includegraphics[width=0.49\textwidth]{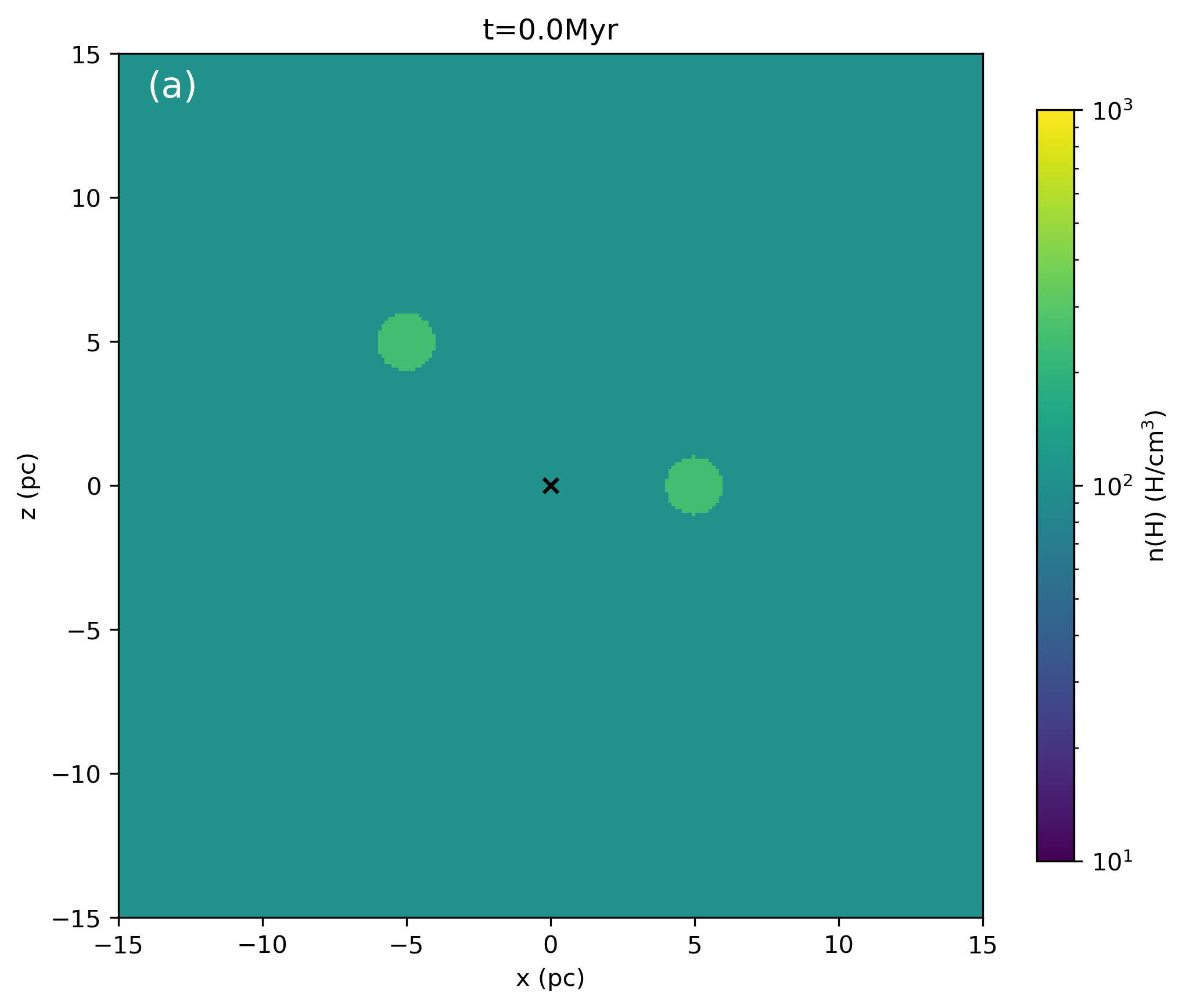} 
        \includegraphics[width=0.49\textwidth]{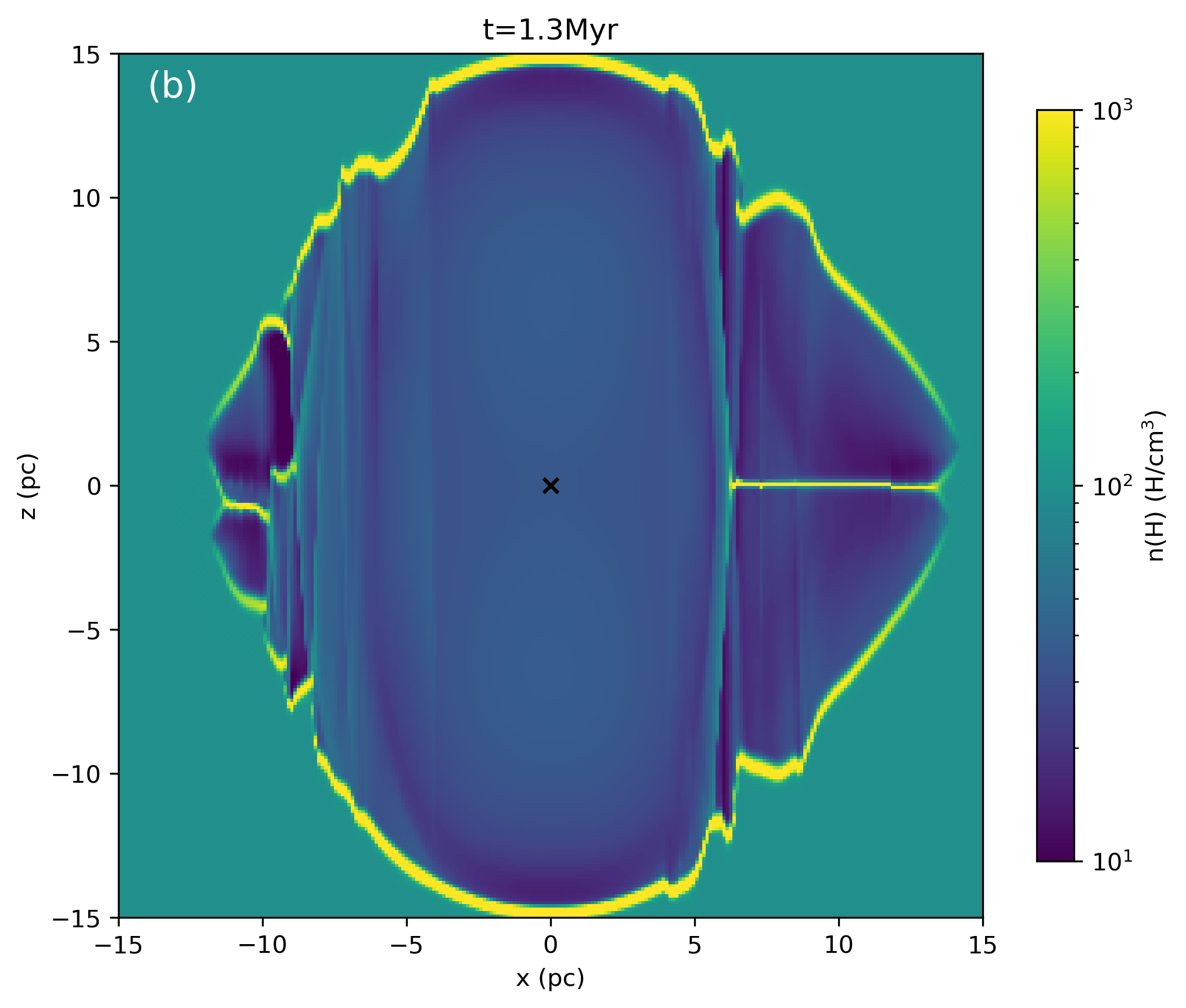}
        \includegraphics[width=0.49\textwidth]{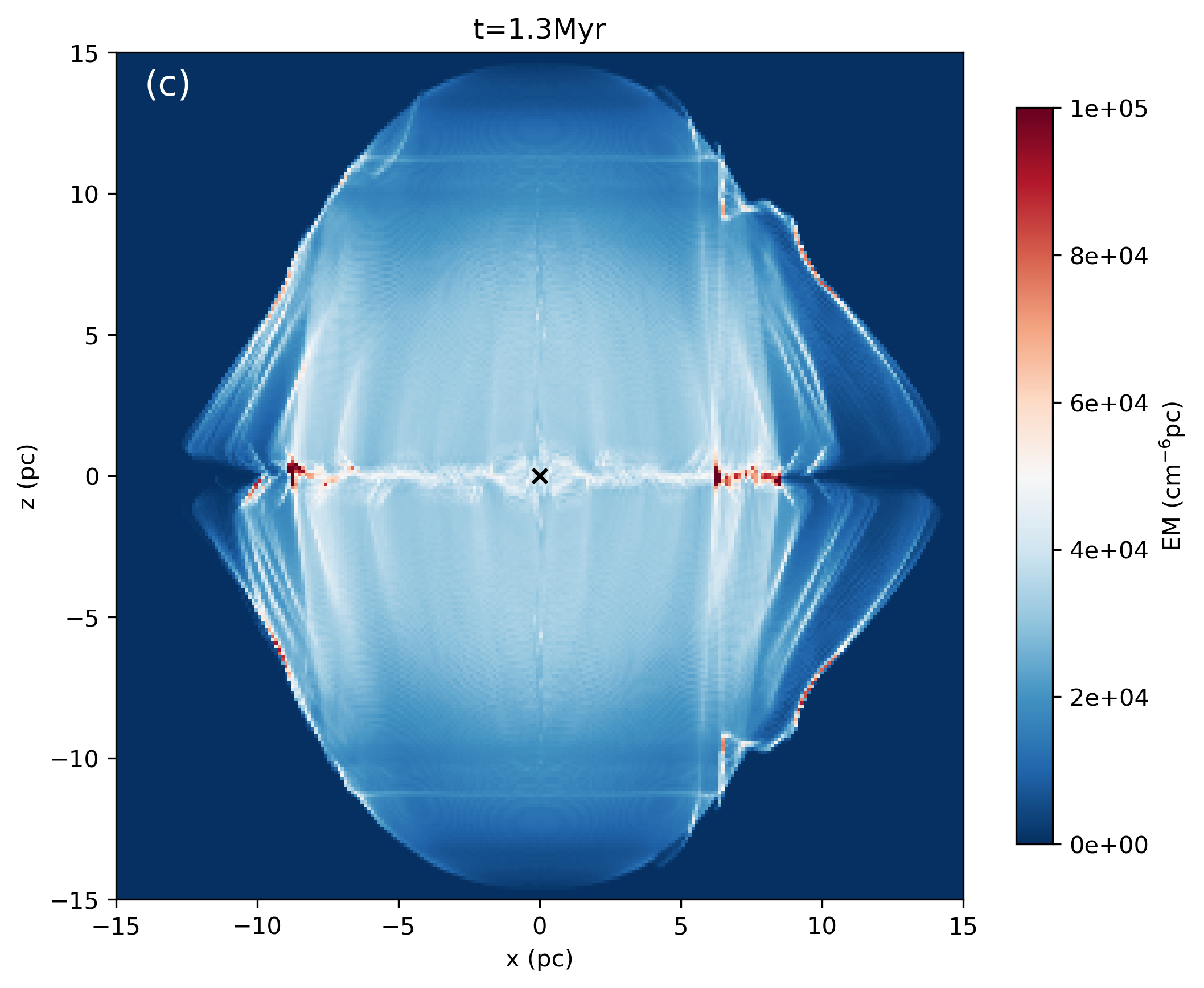}     \includegraphics[width=0.49\textwidth]{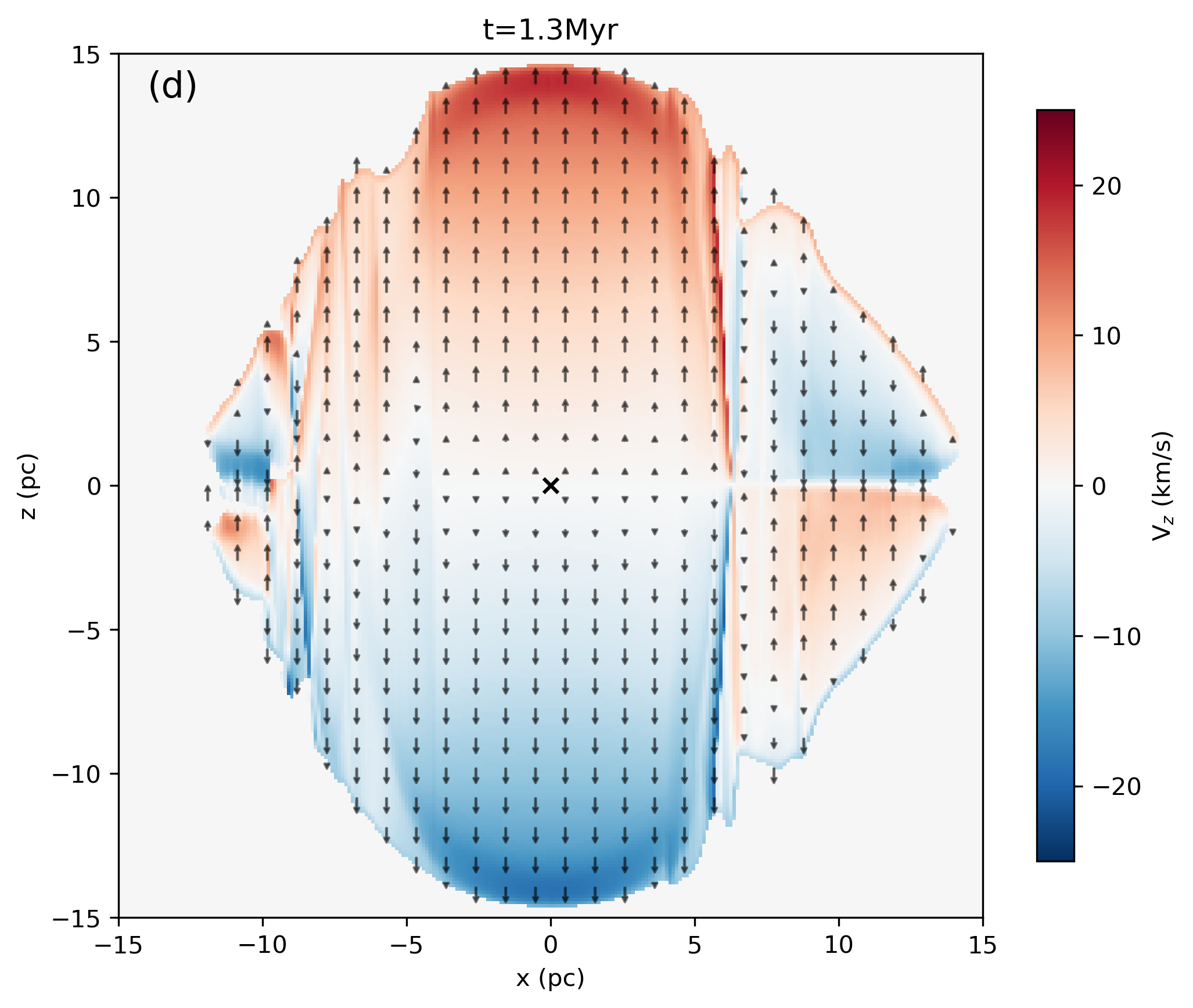}      
        \caption{\label{fig:lowdensblobs} Cutaways from a simulation run with preset blobs of lower density than presented in \S\ref{sec:blob_simul} and Fig. \ref{fig:blob_slices}. Panel (a) shows an $x-z$ slice (along $y=0$) of the number density at $t=0$, i.e. the initial condition before activation of the ionizing source. Panel (b) shows the same at t=1.3 Myr. Panel (c) shows the ionized gas emission measure along the line of sight. Panel (d) shows the the z-velocity, $v_z$, along an $x-z$ slice ($y=0$) with velocity vectors shown in black and scaled to the magnitude of the velocity. The location of the ionizing source is marked with a black cross in all panels.
}
    \end{figure*}

    The results are presented in Figure \ref{fig:lowdensblobs} (similar to Fig. \ref{fig:blob_slices}). Panel (a) shows the number density, n(H), on an $x-z$ slice along $y=0$, at $t=0$, i.e. the initial condition. Panel (b) shows the same at $t=1.3$ Myr. It can be seen that, in contrast to the filament in panel (b) of Fig. \ref{fig:blob_slices}, this filament (density enhancement) displays a curved morphology, which is reinforced by the ionized gas emission measure shown in panel (c) and velocity plot shown in panel (d). This effect can be explained by the lower density of the neutral gas reservoir, which causes a more rapid depletion of neutral gas as the ionization proceeds. Because the filament (ionized gas flow) is sourced at the ``tip'' of the neutral gas reservoir, the rapid erosion of the ``tip'' means that as ionization proceeds, separate, connected channels of ionized gas, at different $x-$values, are created, giving the appearance of a curved filament even though the plasma is still constrained to flow along the local field lines. %Although the filament shown in the figures is much longer than the filaments in Sgr C, 
    A similar mechanism could be invoked to justify the presence of curved filaments in highly magnetized \Hii\:regions like Sgr C. %Such a large spatial scale was adopted here because of resolution constraints.
\end{appendix}

\end{document}